\documentclass[num-refs]{wiley-article}

\usepackage{siunitx}
\usepackage{lineno}

 \usepackage{float} 
   \usepackage{caption}
	\usepackage{subcaption}
	\usepackage{graphicx,rotating,booktabs}
    \usepackage{graphics}
	\usepackage{amsmath}
	\usepackage{pdfpages}
	\usepackage[resetlabels]{multibib}
    \usepackage{mathtools}
    \usepackage{setspace}
    \usepackage{rotating}
\newcites{pp}{References}
\usepackage{lineno,hyperref}
\usepackage[verbose]{placeins}
\usepackage{blindtext}

\papertype{Original Article}

\title{Self-Synchronising On-Off-Keying Visible Light Communication System For Intra and Inter-Vehicle Data Transmission}

\author[1]{Shamin Achari}
\author[2]{Alice Yi Yang}
\author[3]{James Goodhead}
\author[4]{Brendon Swanepoel}
\author[5]{Ling Cheng}

\affil[1,2,3,4,5]{School of Electrical and Information Engineering, University of the Witwatersrand, Johannesburg, Gauteng, 2050, South Africa}

\corraddress{Ling Cheng, School of Electrical and Information Engineering, University of the Witwatersrand, Johannesburg, Gauteng, 2050, South Africa}
\corremail{Ling.Cheng@wits.ac.za}

\fundinginfo{South African National Research Foundation, Grant Number: 112248}

\runningauthor{Achari et al.}

\begin{document}

\maketitle

\begin{abstract}
Visible Light Communication (VLC) is a current technology which allows data to be transmitted by modulating information onto a  light source. It has many advantages over traditional radio frequency communication and up to 10,000 times larger bandwidth. Existing research in visible light communication assumes a synchronised channel, however,  this is not always easily achieved. In this paper, a novel synchronised intra and inter-vehicle VLC system is proposed to ensure reliable communication in both inter and intra-vehicle communication for Infotainment Systems (IS). The protocol achieves synchronisation at the symbol level using the transistor-transistor logic protocol and achieves frame synchronisations with markers. Consequently, the deployment of the protocol in both inter and intra-vehicle communication presents numerous advantages over existing data transmission processes. A practical application, where VLC is used for media streaming is also previewed. In addition, various regions of possible data transmission are determined with the intention to infer forward error correction schemes to ensure reliable communication.

\end{abstract}

\section{Introduction}
\label{sec:intro}
Visible Light Communication (VLC) works on the same principal as optical wireless communication and is composed of three main parts: the transmitter; the channel,or medium, and the receiver. The transmitter modulates data onto the visible light spectrum by means of a Light Emitting Diode (LED) or other light sources. The channel is the free space between the transmitter and receiver. The receiver uses photodiodes or other optical detection devices such as camera lenses to identify the transmitted signals \cite{towardslifi}. This concept is not only limited to the visible light spectrum but is easily extended to the infrared and ultraviolet regions as well. The mainstream idea of VLC is largely brought about due to the advancements in the manufacturing of LEDs as newer LEDs allow for high modulation speeds which provide up to Gbps communication \cite{gbps}.
VLC promises major advantages over traditional Radio Frequency (RF) communication as there is less health risk involved in communication, there is approximately 390 THz of unlicensed bandwidth and there is additional security thanks to the required Line of Sight (LOS) transmission \cite{asurvey, borgia2014internet}. In addition, for applications such as underwater communication, VLC is a strong candidate as RF communication is strongly attenuated by water \cite{asurvey,underwater}. As the Internet of Things (IoT) paradigm makes headway, more subsets of research are explored such as the Internet of Vehicles (IoV), where vehicles will exchange information with other surrounding vehicles, buildings and infrastructure \cite{bazzi2016visible}. VLC will have a high impact in short distance communication in IoT as well as avenues such as IoV and it can be used to improve the performance of traditional vehicular networks when used as a supplementary channel of transmission as shown in \cite{bazzi2016visible}. 

With all the advancements in hardware sensors and processing power, vehicles have seen a rise in the amount of data that is transmitted within itself. From anti-braking systems to stability control, even tire pressure sensing and radar or cameras are being incorporated into vehicles \cite{intra}. All these systems need to transmit data to the main processing unit and as such this increases the wiring requirements and complexities within vehicles. The main form of wireless communication used with many vehicles today is Bluetooth communication where the fundamental use is that of infotainment\cite{intra}. Additionally, since VLC does not work in the RF spectrum, there are many issues of interference in communication devices and instrumentation that is avoided altogether. Hence, VLC is a perfect candidate for intra-vehicle communication.

The majority of research done in the field of VLC tends to focus on achieving faster transmission speeds or producing multiple carrier schemes while few describe synchronisation techniques for such systems. Wang et al achieved speeds of 3.25 Gbps communication using a 512 QAM modulation and single carrier frequency domain equalisation techniques \cite{wang325}. Employing Wavelength Division Multiplexing and Discrete Multi-tone modulation, Vucic et al produced a VLC system which is able to reach average speeds of 803 Mbps across a single LED \cite{803}. Various hardware improvements are employed to gain larger data rates. For example, McKendry et al use special violet $\mu$ LED's and Orthogonal Frequency Division Multiplexing modulation to achieve speeds of up to 3.32 Gbps modulated on a single LED \cite{gbps}. Unique improvements to VLC systems is shown in \cite{cmos} where Chow et al use the rolling shutter effect of a standard CMOS smartphone camera to increase the data rates by up to 60 times to gain a maximum data rate of 1.68 Kbps.

Frame synchronisation in VLC systems is achieved by either adding packet numbers or by using marker codes or extended prefix synchronization codes as shown in \cite{disney}. Comma free codes, number theoretic codes, and watermark codes have all been successfully implemented to protect, either by detection or correction, against synchronisation errors at a symbol level. A more comprehensive discussion of these codes are described in \cite{ling,mercier2010survey}. Alternatively, another layer of synchronisation is obtained at the signal level using techniques such as Phase-Locked Loops, however, this is beyond the consideration of this paper.

Furthermore, VLC has mainly been used as a means of inter-vehicle communication or vehicle to infrastructure communication while little has been said about the possibilities of employing VLC as a communication medium within the constraints of individual vehicles.

The aim of this paper are to introduce a self-synchronising VLC scheme which achieves synchronisation at the symbol level using the Transistor-Transistor Logic (TTL) protocol and achieves frame synchronisations with markers. The novelty and main contribution comes in the form of the synchronisation techniques used in conjunction with the TTL protocol as better speeds can easily be achieved by using better hardware components and circuitry. The system shows the potential to be easily deployed within an automobile or aeroplane which will provide communication within the vehicle. In addition, various regions of possible data transmission are determined with the intention to infer Forward Error Correction (FEC) schemes to ensure reliable communication.

The layout of this paper is as follows: Section \ref{sec:background} discusses the use of an inter-vehicle VLC system, the background of TTL serial communication protocol and channel errors. The system boundaries, limitations and assumptions are given in Section \ref{sec:ttlvlc} along with the components of the proposed system. The results are presented with a detailed analysis using various parameters in Section \ref{sec:results}. Finally the conclusion is given in Section \ref{sec:conclusion}.
\section{Background}
\label{sec:background}
\subsection{TTL Serial Communication Protocol}
\begin{figure}[h]
	\centering
		\centering
		\includegraphics[width=9.5cm]{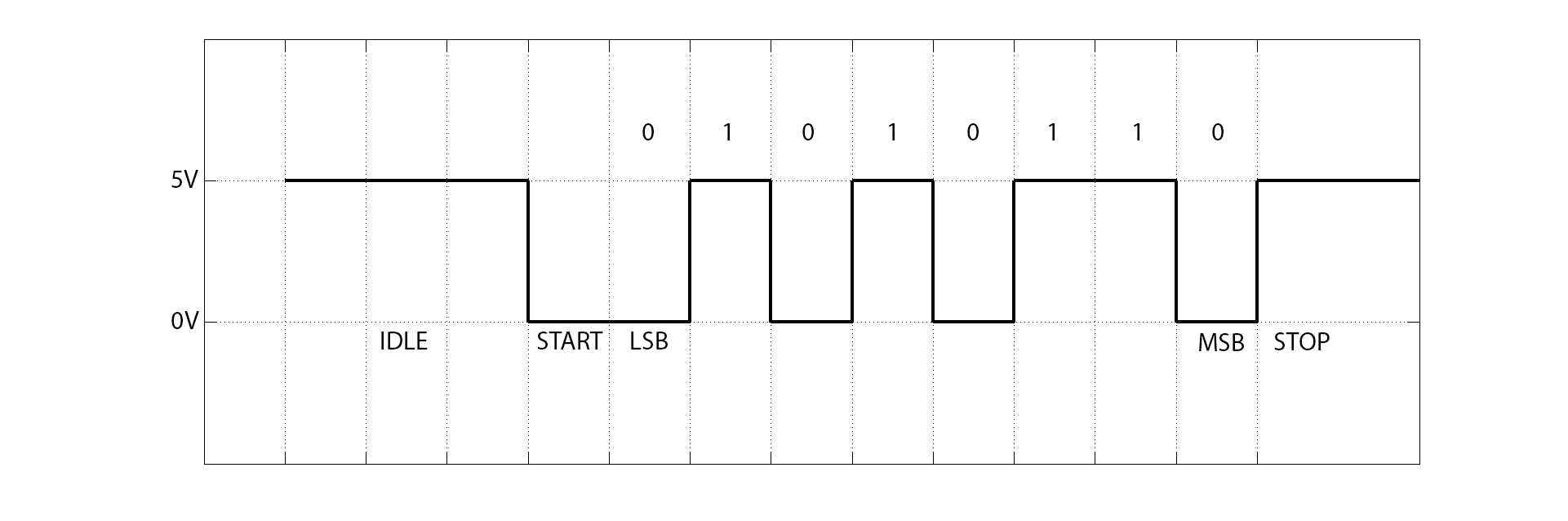}
		\caption{Example bit stream when communicating with TTL serial protocol}\label{fig:uart}		
	\end{figure}

The TTL serial communication protocol provides asynchronous communication and thus no clock signal is transmitted. The communication speed and other parameters such as parity and number of data bits need to be agreed upon beforehand by both the transmitter and receiver. The TTL serial protocol encompasses its data bits (7 or 8 bits) between a start bit and by either one or two stop bits. Data is transmitted with the least significant bit first and lastly, parity bits (even, odd, mark or space) may be added if desired. In TTL Serial communication a 1 is represented as either 3.3V or 5V depending on the hardware device used while a 0 is represented by 0V. Figure \ref{fig:uart} shows an example bitstream when the number 106 (0110 1010$_B$) is transmitted with no parity bits using the TTL serial protocol. It is this TTL serial protocol with its start and stop bits that introduce the synchronisation errors.

\subsection{On-Off-Keying (OOK) Modulation Scheme}
Modulation schemes allow information to be encapsulated in a certain manner in order to be transmitted. On-Off-Keying (OOK), Pulse Position Modulation (PPM), Inverse Pulse Position Modulation (IPPM) and Variable Pulse Position Modulation (VPPM) are all examples of time domain modulation schemes \cite{shlomi}. OOK is the simplest method of modulation and is a form of binary signal modulation. In VLC-OOK a 1 is represented by turning on an LED for a period of, \textit{T} seconds, whereas a 0 is represented by switching the LED off for \textit{T} seconds \cite{shlomi}. OOK modulation is used in this set-up as the TTL protocol signal can be easily modulated onto the LED with no additional processing.

\subsection{Synchronisation Errors} 
The errors that occur in asynchronous communication include insertion, deletion and substitution errors where insertion and deletion errors make up synchronisation errors \cite{ling}. Insertion errors could occur when the receiver samples too fast (over-samples) and essentially add more symbols to the received data stream whereas, deletion errors occur when the receiver samples too slow (under-samples) and remove symbols from the transmitted data stream \cite{ling}. Substitution errors occur when a received symbol is transformed to another symbol of that alphabet \cite{ling}.

\subsection{Inter and Intra-Vehicle Communications}
Inter-vehicle communications between vehicles on highways are generally used to increase road safety and reduce traffic congestion \cite{821160-traffic}. It has been further employed to assist human operators, by providing communication between the sensors and systems for synchronized operations \cite{6819448-intra-vehicle-communication}. The traditional techniques for intra-vehicle communication are performed over Control Area Network (CAN) and FlexRay networks for Electronic Control Units (ECUs) installed into vehicles \cite{6819448-intra-vehicle-communication, specification1991robert, makowitz2006flexray}. The traditional techniques proved reliable for the communication of bus-systems. However, it presented problems through the number of ECUs required to connect to the system over time. Consequently, the increase in ECUs and connections meant that it monopolised a significant amount of bandwidth. Thus, rendering the architecture of the traditional techniques as inefficient for modern networks, particularly for the Driver Assistance System (ADS).

The most up-to-date intra-vehicle communication is performed over an Ethernet network. In the Ethernet network, the physical layer is crucial as it is the physical material employed to transmit information to and from a centralised location. In the physical layer, there are two standards: automotive-specific and non-automotive specific. The automotive-specific standards are specifically tailored for intra-vehicle communications which are used to transmit information between vehicles to control traffic \cite{kopetz1993ttp}. Popular automotive-specific standards are CAN, FlexRay, and Local Interconnect Network (LIN). For non-automotive specific, popular standards are Low-Voltage Differential Signaling (LVDS), which is constructed with twisted pair copper cable for high-speed signalling transmission. Alternatively, Firewire is another commonly utilised non-automotive specific standard, in which it communicates through the use of video cameras in computer communication bus standards \cite{rabel2001integrating, ke2013survey}.

However, with the various number of sensors utilised for vehicle assistance operation, the number of wired connections between sensors and systems become cumbersome and requires space to be reserved for wire placements. Consequently, the number of wires add to the overall weight of the vehicle. Narmanlioglu, O Et. al. \cite{NARMANLIOGLU2018138} conducted an investigation in the viability of VLC in inter-vehicular connectivity for both point-to-point vehicular VLC and decode-and-forward relaying based cooperative vehicular VLC. This includes relay terminals between the source and destination terminals, in order to ensure road safety. The investigation considered direct current biased optical orthogonal frequency division multiplexing (DCO-OFDM) based MIMO transmission scheme. The authors evaluated the performance using different MIMO modes. The different modes include Repetition Code (RC) and Spatial Multiplexing (SM). The performance was further evaluated using different modulation orders with different transmitters. The results obtained by the authors indicated that transmitters and receivers closest to one another achieved better performances. This is due to the high signal-to-noise ratio requirement for RC modes. Whilst the SM performed poorly due to the channel correlation. The authors concluded with the SM mode being more favourable for MIMO, as RC requires higher order modulation.

In addition to inter and intra-vehicle communication, Wireless Avionics Intra-Communications (WAIC) is introduced for the communication between single points within aircraft \cite{torres2016enabling}. WAIC utilises radio communication for the transmission of data between two points. The purpose of WAIC is to reduce the redundancy of wired connections within the aircraft. The reduction of wired connections reduces the overall weight of the aircraft, as wired connections for modern aeroplanes are approximately 500 kilometres in length with an estimated weight of 7,400 kilograms for the Boeing 787 \cite{war_2017}. As a result, a wireless system will increase fuel efficiency and simplify the wiring within the aircraft. 

\section{An Original Intra and Inter-Vehicle VLC System Synchronised With the Transistor to Transistor Logic Protocol and Markers}
\label{sec:ttlvlc}
\begin{figure}[h]
	\centering
		\centering
		\includegraphics[width=10cm]{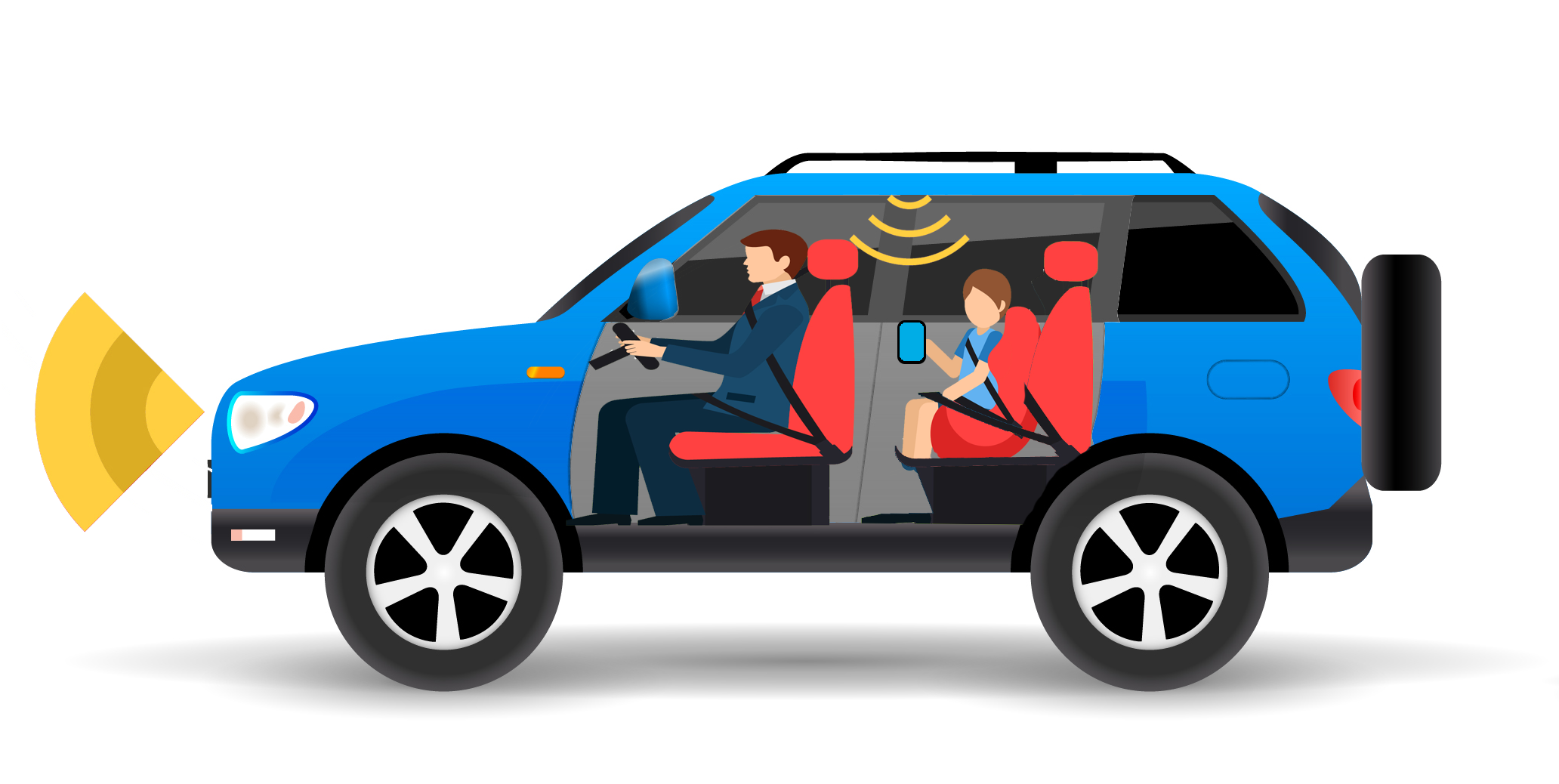}
		\caption{VLC system for both inter and intra-vehicle communication\protect\footnotemark}\label{fig:car}	
	\end{figure}

\begin{figure}
    \centering
    \begin{subfigure}[b]{12.5cm}            
            \includegraphics[width=12.5cm]{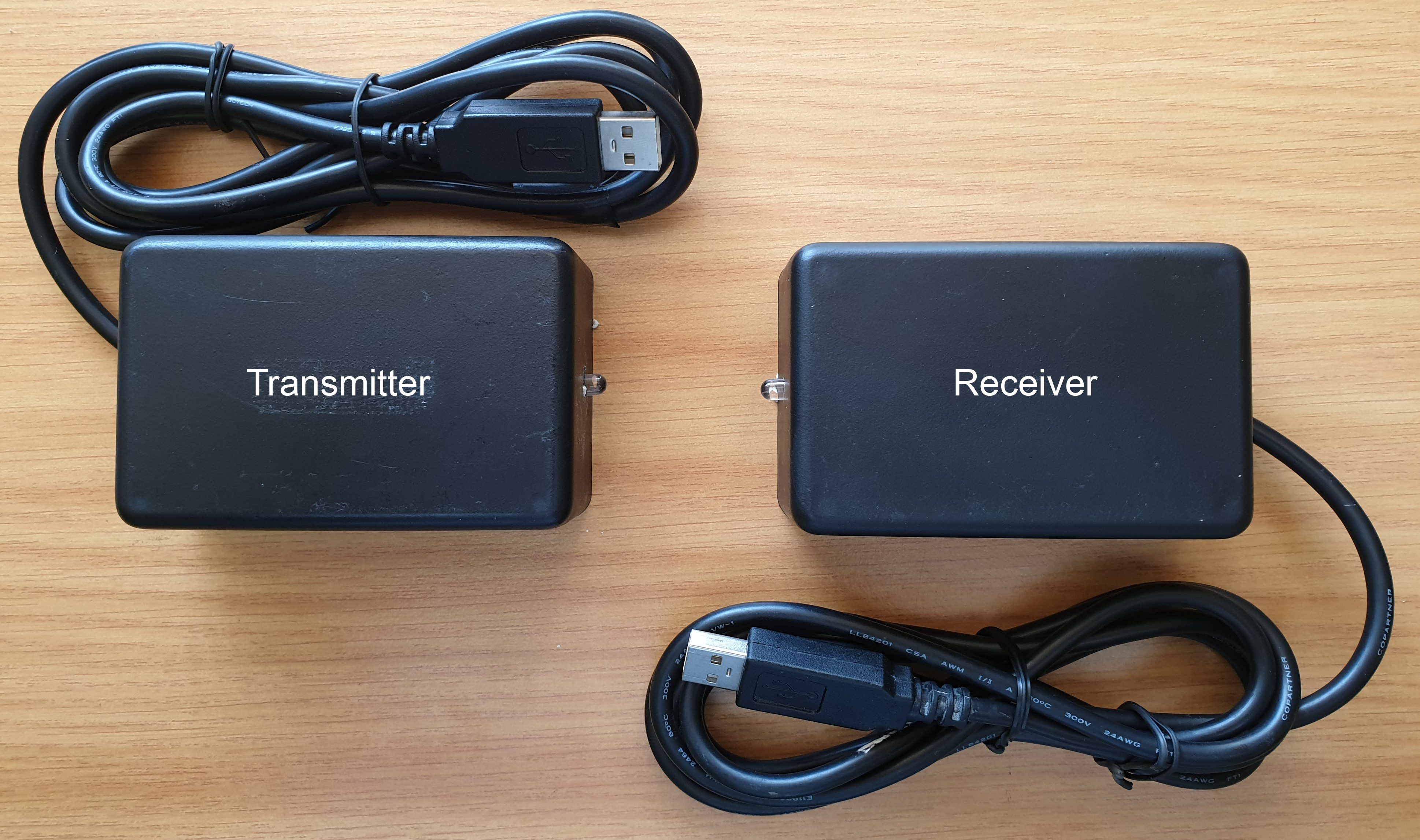}
            \caption{Housed transmitter and receiver components}
            \label{fig:txandrx}
    \end{subfigure}
    \begin{subfigure}[b]{6.2cm}            
            \includegraphics[width=6cm]{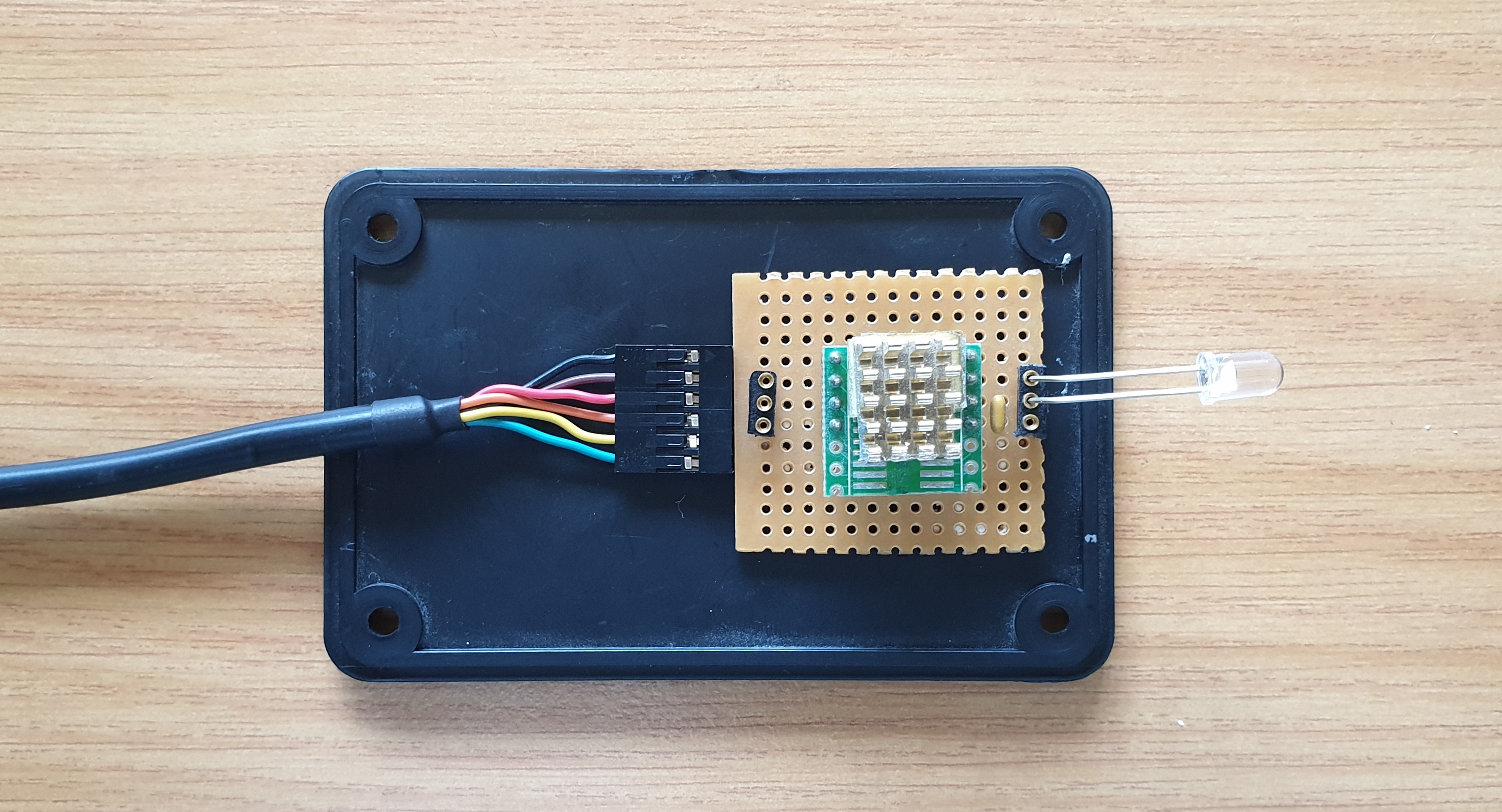}
            \caption{Detailed view of the transmitter components}
            \label{fig:txImage}
    \end{subfigure}
    \begin{subfigure}[b]{6.5cm}
            \centering
            \includegraphics[width=6cm]{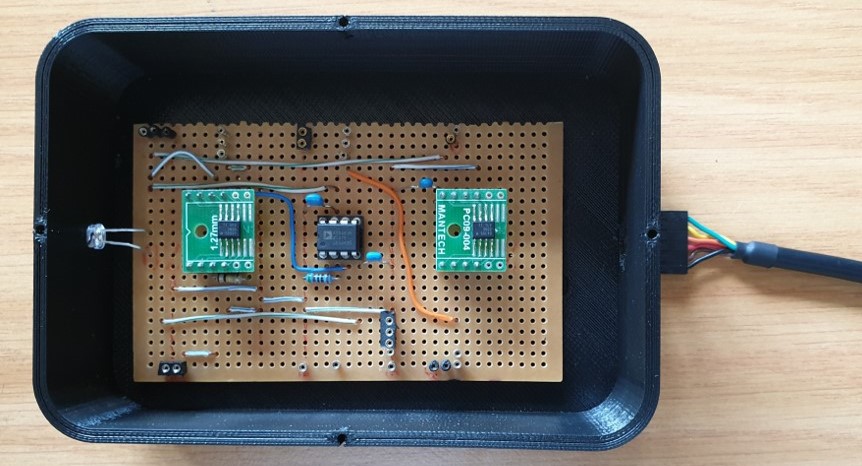}
            \caption{Detailed  view of the receiver components}
            \label{fig:rxImage}
    \end{subfigure}
    \caption{Electronic Components of the VLC system with its casing}\label{fig:PhysicalVLCSystem}
\end{figure}

A new VLC system for wireless communication within a single vehicle and between vehicles is illustrated in Figure \ref{fig:car}, in which existing light sources are employed for inter and intra-vehicle communication. This setup, which uses the TTL serial transmission protocol and OOK to asynchronously transmit data, is implemented and tested. This system is illustrated in Figure \ref{fig:PhysicalVLCSystem}. The transmitter and receiver with its housing is shown in Figure \ref{fig:txandrx}. Figures \ref{fig:txImage} and \ref{fig:rxImage} present the intricacies inside both transmitter and receiver, respectively.

	\footnotetext{Image adapted from Freepik.com, artwork originally desigend by macrovector}

\subsection{Use in Vehicular Communication}
\begin{figure}[h]
	\centering
		\centering
		\includegraphics[scale=0.6]{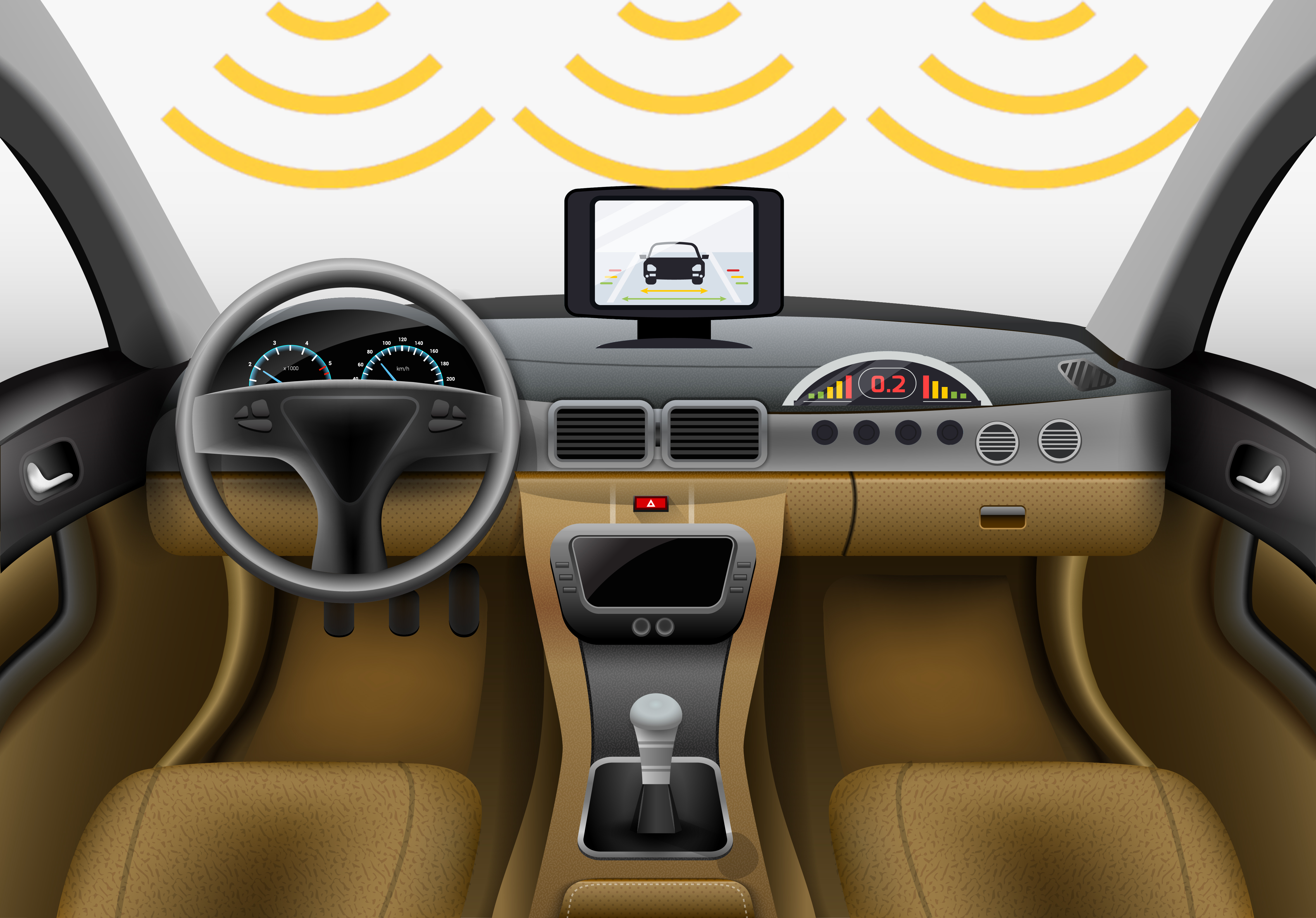}
		\caption{VLC system for intra-vehicle communication\protect\footnotemark}\label{fig:carinside}		
	\end{figure}

The compact size and ease of access to information of this VLC system allow it to be easily deployed within any type of vehicle to provide a means of data communication. The main benefit of using a VLC system for intra-vehicle data transmission is the fact that it reduces the overhead and complexity that its wired counterpart shares. The implementation concept of the VLC system for intra-vehicle communication is shown in Figure \ref{fig:carinside}. Intra-vehicle communication has the potential to be extended to aeroplanes as Wireless Avionics Intra-Communication (WAIC) to simplify wired connections. The implementation of this system largely reduces the overall weight of the aircraft. Traditionally, wired connections for a Boeing 787 aircraft requires an estimated length of 500 kilometres of cabling \cite{war_2017, torres2016enabling}. The system employs the same concept as intra-vehicle communication and has the ability to work without the use of visible light by simply utilising the infrared spectrum and infrared LEDs.

	\footnotetext{Image adapted from Freepik.com, artwork originally desigend by macrovector}

\subsection{Transmitter}
\begin{figure}[h]
	\centering
		\centering
		\includegraphics[width=7cm]{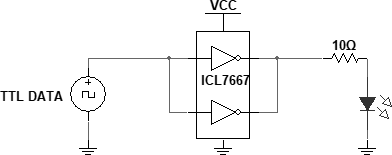}
		\caption{Circuit diagram of VLC transmitter circuit using a MOSFET driver}\label{fig:transbd}		
	\end{figure}

The transmitter, whose circuit diagram is shown in Figure \ref{fig:transbd}, is responsible for transmitting the TTL serial data. It consists of a USB to serial cable as this allows for easy connection to a laptop where the data is generated. This data is then propagated to an analogue circuit that contains an LED which modulates the data using OOK. A MOSFET or LED driver is used as the switching device as it allows for high data rates while supplying the required current for a high powered LED.

\subsection{Receiver}
\begin{figure}[h]
	\centering
		\centering
		\includegraphics[width=7cm]{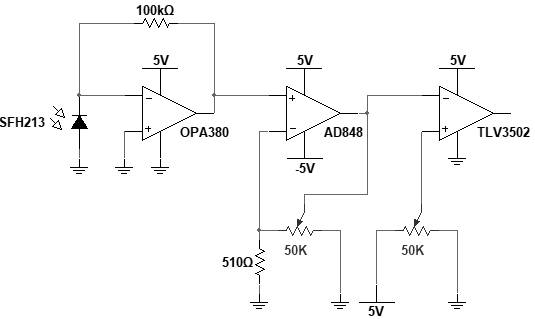}
		\caption{Circuit diagram of VLC receiver circuit}\label{fig:recbd}		
	\end{figure}

The receiver is responsible for capturing the variations in light intensity, determining a binary stream from these fluctuations and then transmitting this binary stream to a laptop or PC to be processed at a later stage. A PIN photodiode is used to capture the light intensity and consequently, it produces a current which is proportional to the intensity of the light that falls on it. A Transimpedance Amplifier (TIA) is then used to convert the current produced, by the photodiode, to a voltage. Preliminary tests proved that using a single TIA to both convert current and amplify the signal is not adequate as a trade-off between the gain and bandwidth is seen in the op-amp. For this reason two separate steps, one for capturing the light signal and the other for amplifying the signal, is used. As an additional measure, the signal will need to be conditioned using a comparator to ensure that a perfect representation of the TTL data is produced. The receiver circuit makes use of the SFH213 photodiode\cite{sfh}, along with the OPA380 and a 100K$\Omega$ feedback resistor which is shown in Figure \ref{fig:recbd}. The combination of these components provides a maximum transimpedance bandwidth of approximately 3MHz as the SFH213 has an internal capacitance of 11 pF. Additionally, the SFH213  has a large detection wavelength range of 400 nm to 1100 nm. This indicates that any colour LED, including white, may be detected.     

\subsection{Test Procedure}
\begin{figure}	
	\centering
	\begin{subfigure}[t]{1.5in}
		\centering
		\includegraphics[width=1.5in]{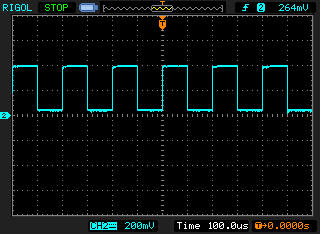}
		\caption{Output of TIA at 10 Kbps}\label{fig:10k}		
	\end{subfigure}
	\quad
	\begin{subfigure}[t]{1.5in}
		\centering
		\includegraphics[width=1.5in]{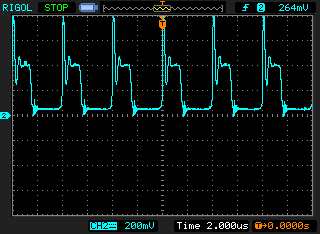}
		\caption{Output of TIA at 100 Kbps}\label{fig:100k}	
	\end{subfigure}
	\quad
    	\begin{subfigure}[t]{1.5in}
		\centering
		\includegraphics[width=1.5in]{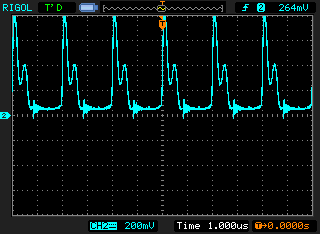}
		\caption{Output of TIA at 1 Mbps}\label{fig:1m}	
	\end{subfigure}
	\caption{Waveforms for different baud rates showing distortions}\label{fig:distort}	
\end{figure}

The transmitter and receiver are placed in a covered, black, polystyrene box at a distance of 15cm from each other. The distance of 15cm ensures no reflections and multipath signals are detected by the photodiode. Any distance exceeding approximately 20cm starts to introduce multipath reflections. This phenomenon is further explained in Section \ref{section:losnlos}. It is assumed that reflections are negligible in the practical system as the system will operate ideally in open free space where reflections are minimised. In addition, the use of the covered box is to ensure consistency of the parameters during tests. This allows the tests to be conducted at any time of the day without the issue of varying ambient light. The system, however, still remains operational in open free space without the box. Section \ref{section:losnlos} describes the effects of Line of Sight (LOS) and Non-Line of Sight (NLOS) further.
Various Signal to Noise Ratio (SNR) values are achieved by adjusting the power of the transmitter which is analogous to increasing or decreasing the distance between the transmitter and receiver. The SNR values used are an approximation and discussed further in Appendix \ref{app:snr}. The SNR values used are evaluated at a baud rate of 10Kbps to ensure little to no distortion is present at the output of the TIA. Figure \ref{fig:distort} demonstrates the distortion of signals at the TIA for various baud rates. A combination of parameters are tested at each SNR value, these include baud rates (10Kbps, 50Kbps, 100Kbps, 250Kbps, 500Kbps, 750Kbps, 1Mbps), frame length (1003 symbols, 5003 symbols, 10003 symbols) and synchronisation symbol length (1 symbol, 5 symbols, 10 symbols).

The baud rates are chosen so that they span an effective range while not exceeding the hardware capacity as the USB to TTL only allows a maximum of 3Mbps. The payload for the frame is a pseudo-random generated string of symbols whose ASCII  values fall in the range of 64 to 126 (@ to \textasciitilde). Each frame contains the exact same payload to allow for an easier analysis, however, a frame ID of three symbols is added to the front of each packet. There is a synchronisation word that is fixed to the beginning of each frame, in this case, it is a varying number of \$ symbols. This synchronisation symbol does not appear in the payload.

A maximum time of 60 minutes, for transmission for a given set of parameters is used. If no substitution errors occur within the first 15 minutes of transmission or no packets are dropped, the transmission ends as the communication is assumed to be in a perfect channel with no appreciable errors and a new set of parameters are chosen. If there are at least 100 packets with substitution errors or 100 packets are dropped, the transmission also ends as this provides enough statistics for a given set of parameters. Any synchronisation errors (when a received frame length is longer or shorter than the expected length) are analysed as a dropped packet while a missed frame is also catagorised as a dropped packet. The frequency of substitution errors is recorded to infer a viable error correction scheme for a given set of parameters.

\subsection{Line of Sight and Non Line of Sight Detection}
\label{section:losnlos}
LOS is defined as a direct, uninterrupted link between the transmitter and receiver whereas a NLOS  uses diffuse reflections off surfaces to form an indirect link between the transmitting LED and the photodiode \cite{kahn1997wireless}. Using LOS increases the power efficiency of the channel while also reducing distortion from multipath signals, whereas, using NLOS increases the robustness and practicality of the system as no direct link is required \cite{kahn1997wireless}.
Figure \ref{fig:losnlos} in Appendix \ref{app:snr} illustrates the LOS and NLOS links that are used to transmit data via the LED to the photodiode. The channel gain for both LOS, $H_{LOS}$, and NLOS, $H_{LOS}$, in a first order reflection scenario can be calculated using the generalised Lambertian radiant intensity and is shown in Equation \eqref{eqn:los} and Equation \eqref{eqn:nlos} in Appendix \ref{app:snr} respectively. This channel gain can essentially be used to determine the total power detected by the photodiode for both direct and reflected links by multiplying it by the transmit power.

\begin{figure}[h]
	\centering
		\centering
		\includegraphics[width=7cm]{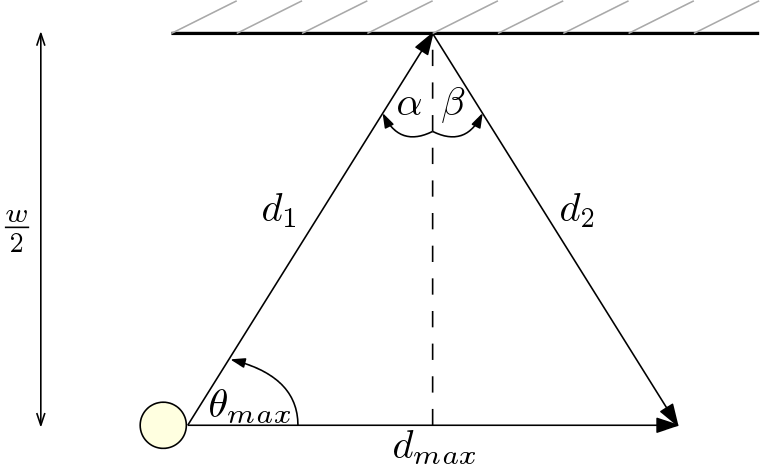}
		\caption{Geometry of testbed showing maximum LOS only distance}\label{fig:boxgeo}		
	\end{figure}

The underlying TTL protocol relies heavily on timing and thus sampling at the correct instance is of the utmost importance. NLOS signals are therefore minimised for the test procedure, as the NLOS signals are superimposed onto the LOS signal which alters the overall received signal.
As stated previously, removing the NLOS signal replicates the real world application scenario as the transmitter and receiver pair will not be in close proximity to surfaces which cause the NLOS links. Figure \ref{fig:boxgeo} presents the geometry of the testbed as well as the maximum allowed distance between the transmitter and receiver to ensure that only LOS links are detected. The setup has a width, $w$, of roughly $35cm$ and uses an LED with a full viewing angle of $120^\circ$. This means the maximum angle of incidence for the light rays, $\theta_{max}$ is $60^\circ$. After using the trigonometry and dimensions of the setup, it is found that $d_1$ is equivalent to $d_2$ and $d_{max}$ due to symmetry. Additionally, $\alpha + \beta$ is $60^\circ$ in this case. This ultimately leads to a maximum distance, $d_{max}$ of roughly $20cm$ between the transmitter and receiver to allow only LOS signals to be detected. Equation \eqref{eqn:dmax} describes a more generalised solution to calculate the maximum distance for only LOS signals to be detected.


\begin{equation}
\label{eqn:dmax}
	\begin{aligned}
		d_{max} = \frac{\frac{w}{2} sin(\alpha+\beta)}{sin^2(\theta_{max})}
	\end{aligned}
\end{equation}

\section{Results and Analysis}
\label{sec:results}
\begin{figure}[h]
	\centering
		\centering
		\includegraphics[width=15cm]{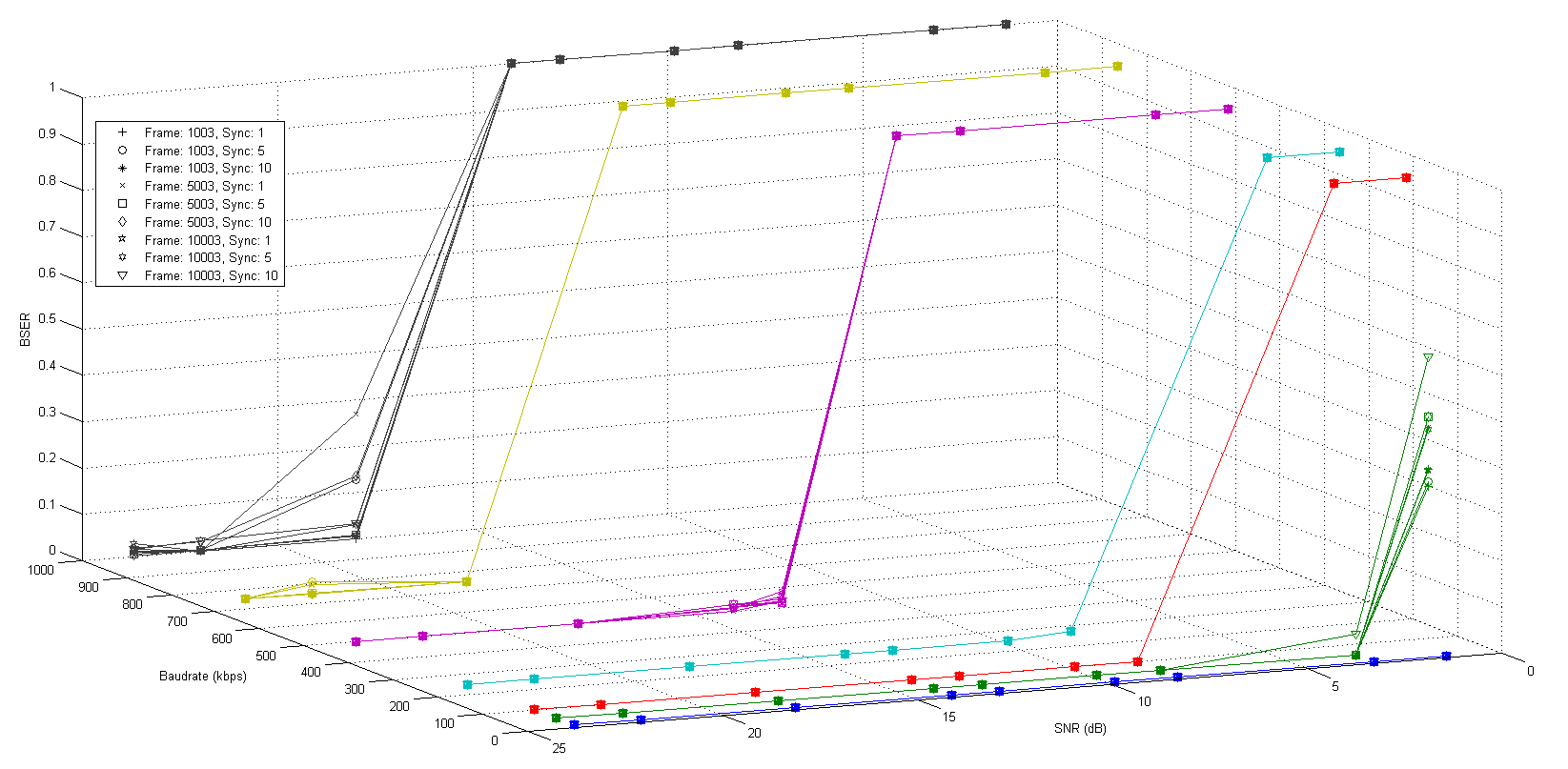}
		\caption{$p_{BSE}$ for various SNR, baud rates, frame lengths and synchronisation lengths}\label{fig:all3d}		
	\end{figure}

Figure \ref{fig:all3d} shows the block synchronisation error rate, $p_{BSE}$, obtained for various SNR and baud rates, as well as different parameters of frame and synchronisation lengths. $p_{BSE}$ is defined as the ratio of blocks that lose synchronisation and the total number of transmitted blocks. A block synchronisation error rate of 0 shows perfectly synchronised communication as all blocks or packets are received without synchronisation errors. A block synchronisation error rate of 1 implies total loss of all packets transmitted. The inverse of the $p_{BSE}$ is defined as reliability which  implies how well data transmission may occur without introducing synchronisation errors. Reliability is characterised as $(1-p_{BSE})$. In Figure \ref{fig:all3d}, most of the variations of parameters perform equally and tend to overlap with major discrepancies only occurring at high baud rates or low SNR values. Additionally, it is evident that synchronisation errors are either catastrophic or cause very little dropped packets that are not in a viable error range to be corrected. In this case, an Automatic Repeat Request (ARQ) system is better suited for synchronisation errors. Regions where data transmission is viable will still benefit from substitution \color{black} FEC \color{black} schemes and these schemes can be deduced from the error frequency plots, as well as reliability plots.

\subsection{Effects of Synchronisation Length}
\begin{figure}	[th]
	\centering
	\begin{subfigure}[]{4cm}
		\centering
		\includegraphics[width=4cm]{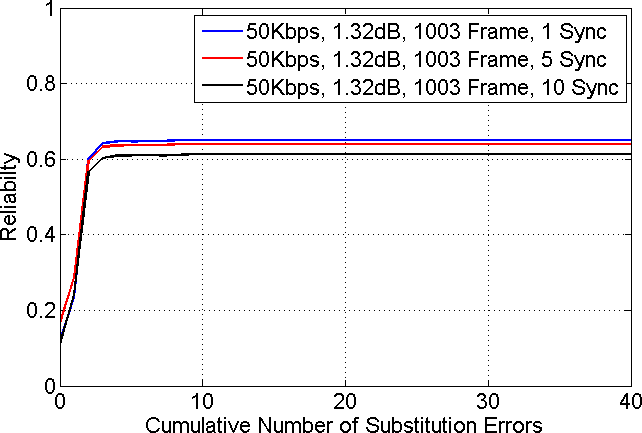}
		\caption{Frame Length:1003}\label{fig:cdf1000f}		
	\end{subfigure}
	\quad
	\begin{subfigure}[]{4cm}
		\centering
		\includegraphics[width=4cm]{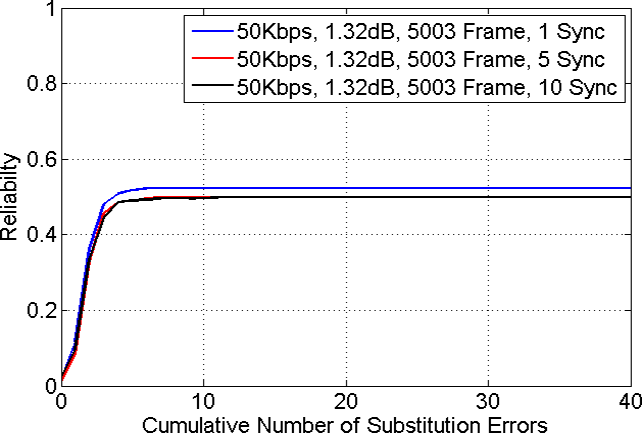}
		\caption{Frame Length:5003}\label{fig:cdf5000f}	
	\end{subfigure}
    	\begin{subfigure}[]{4cm}
		\centering
		\includegraphics[width=4cm]{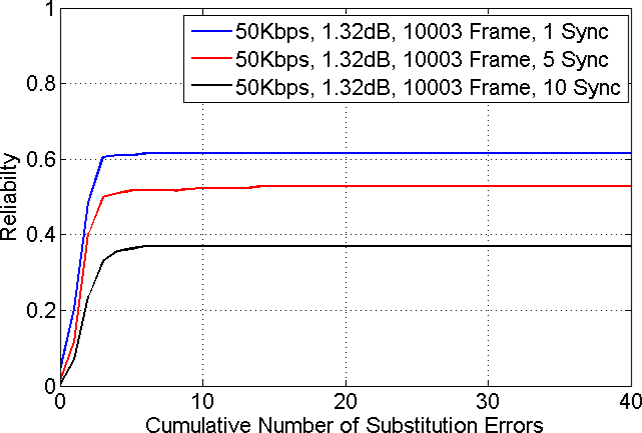}
		\caption{Frame Length:10003}\label{fig:cdf10000f}
	\end{subfigure}
	\caption{CDF indicating reliability using various sync lengths at 50Kbps and 1.32dB}\label{fig:cdfsync}	
\end{figure}

\begin{figure}[h]	
	\centering
	\begin{subfigure}[]{4cm}
		\centering
		\includegraphics[width=4cm]{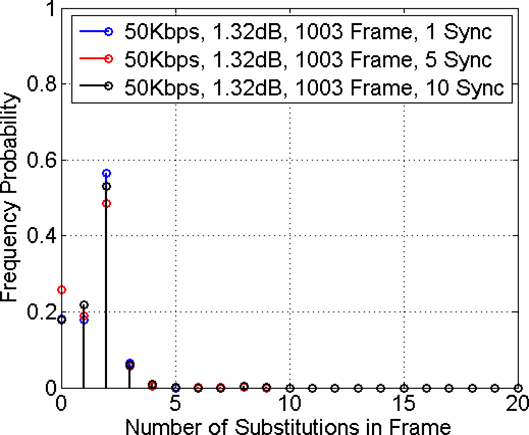}
		\caption{Frame Length:1003}\label{fig:pdf1000f}		
	\end{subfigure}
	\quad
	\begin{subfigure}[]{4cm}
		\centering
		\includegraphics[width=4cm]{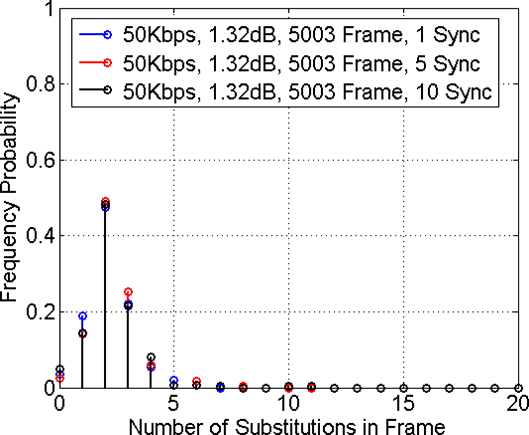}
		\caption{Frame Length:5003}\label{fig:pdf5000f}	
	\end{subfigure}
    	\begin{subfigure}[]{4cm}
		\centering
		\includegraphics[width=4cm]{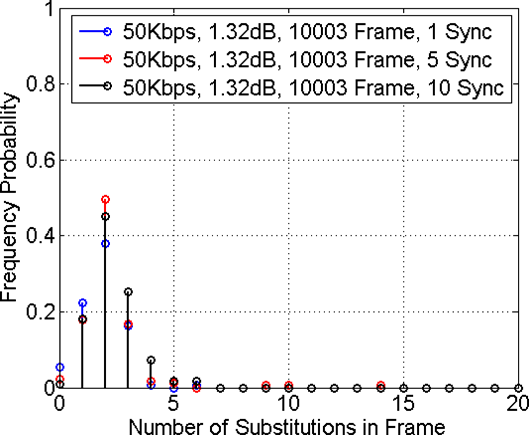}
		\caption{Frame Length:10003}\label{fig:pdf10000f}
	\end{subfigure}
	\caption{PDF indicating frequency of substitution errors using various sync lengths at 50Kbps and 1.32dB}\label{fig:pdfsync}	
\end{figure}

These results show the effects of changing the length of the synchronisation symbols with respect to the reliability as well as substitution errors in data transmission. From Figure \ref{fig:cdf1000f} it can easily be seen that a synchronisation word of one symbol has the least amount of dropped packets followed by five synchronisation symbols and lastly ten. This is due to the fact that there is a lower probability associated with missing fewer consecutive symbols. As the synchronisation symbol length increases, more packets are dropped as there is a higher probability of missing a synchronisation symbol, however, the packets that are received at higher synchronisation word lengths are more likely to be perfect or have very few substitution errors. This phenomenon can be seen in Figure \ref{fig:pdf1000f}. This trend continues at higher frame lengths too and is shown for a frame length of 5000 in Figure \ref{fig:cdf5000f} and Figure \ref{fig:pdf5000f} as well as a frame length of 10000 shown in Figure \ref{fig:cdf10000f} and Figure \ref{fig:pdf10000f} which shows the Cumulative Density Function (CDF) and Probability Density Functions (PDF) respectively. From Equation \eqref{eqn:pfail} and Equation \eqref{eqn:perr} in \ref{app:probs}, it can be seen that an increase in synchronisation symbols increases the likelihood of synchronisation failure while decreasing the chance of synchronisation error.

\subsection{Effects of Frame Length}
\begin{figure}[ht]	
	\centering
	\begin{subfigure}[]{6cm}
		\centering
		\includegraphics[width=6cm]{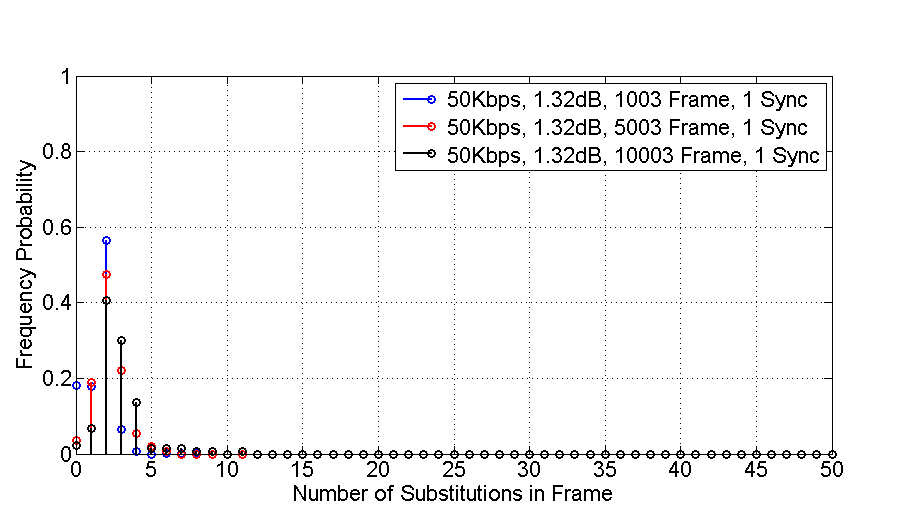}
		\caption{PDF for various frame lengths}\label{fig:pdf1s}		
	\end{subfigure}
	\quad
    	\begin{subfigure}[]{6cm}
		\centering
		\includegraphics[width=6cm]{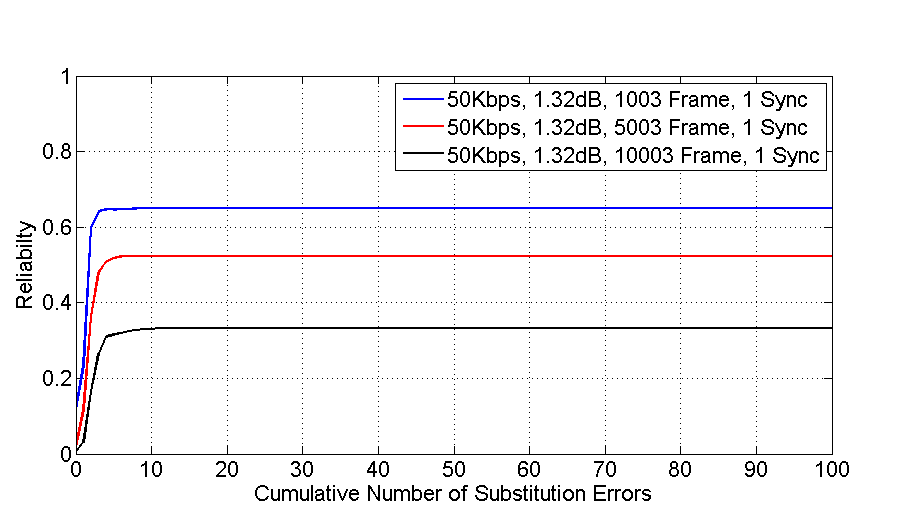}
		\caption{CDF for various frame lengths}\label{fig:cdf1s}
	\end{subfigure}
	\caption{PDF and CDF showing reliability using the various frame lengths at 50Kbps and 1.32dB}\label{fig:pdfsync}	
\end{figure}

The following results show the effects of different frame lengths on the number of errors produced. The tests are performed at an SNR of 1.32dB, 50Kbps baud rate with a synchronisation word length of 1 symbol. Frame lengths of 1003, 5003 and 10003 are tested of which the error frequency is shown in Figure \ref{fig:pdf1s} and the reliability is shown in Figure \ref{fig:cdf1s}. From these figures it is evident that shorter frame lengths outperform their longer counter parts in both the dropped packet frequency, as well as having more packets with less substitution errors. This intuitively makes sense and can be illustrated using an example. If the probability of causing an error (either insertion, deletion or substitution) is for instance 0.01\% per symbol transmitted. We then have a chance of producing approximately 10 errors for the frame of 1003, 50 errors for a frame of 5003 and 100 errors for a frame of 10003. So as the frame length increases, the chances of affecting more symbols also increases and thus more errors are introduced. From Equation \eqref{eqn:pfail} and Equation \eqref{eqn:perr} in Appendix \ref{app:probs}, it can also be shown that increasing the frame length effectively increases synchronisation failure and error rates. It is worth noting that the substitution errors caused by $p_{err}$ are not the same as the the Symbol Error Rate (SER).

 \subsection{Effects of high SNR}
\begin{figure}	
	\centering
	\begin{subfigure}[t]{6.5cm}
		\centering
		\includegraphics[width=6.5cm]{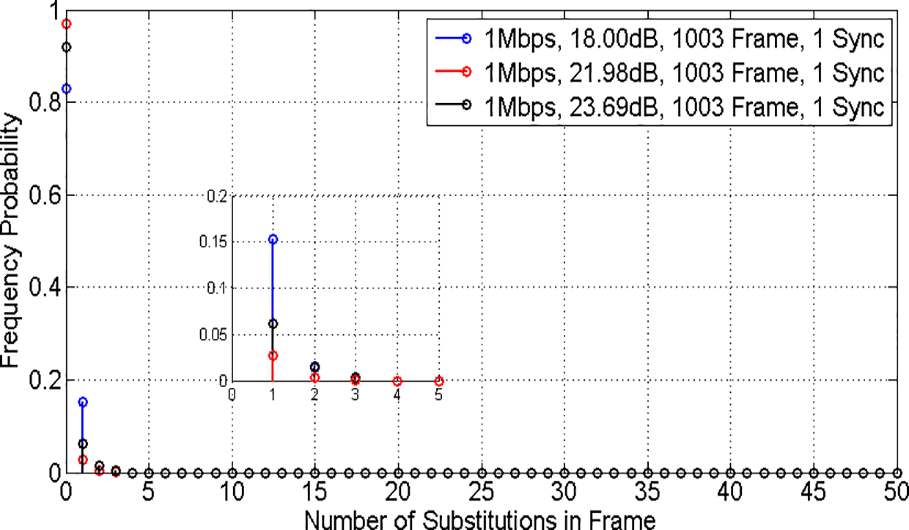}
		\caption{PDF for various SNR values}\label{fig:pdf1M}		
	\end{subfigure}
	\quad
    	\begin{subfigure}[t]{6.5cm}
		\centering
		\includegraphics[width=6.5cm]{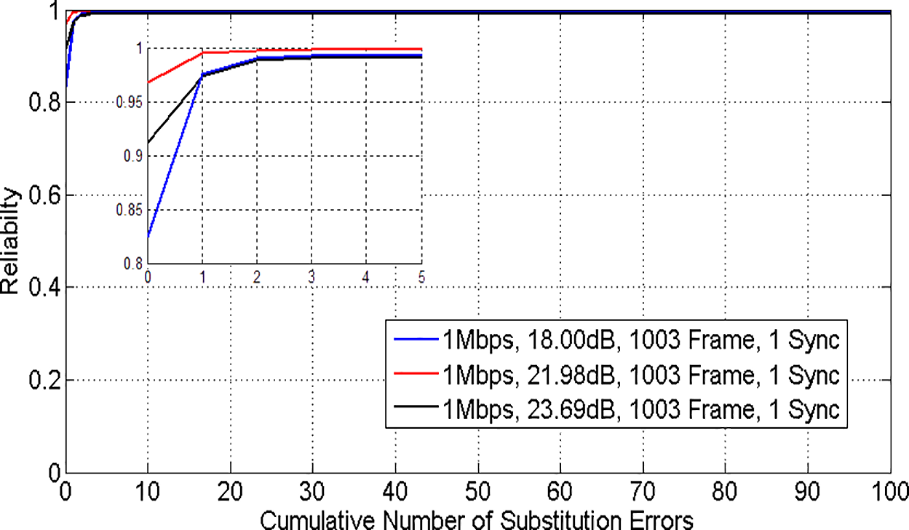}
		\caption{CDF for various SNR values}\label{fig:cdf1M}
	\end{subfigure}
	\caption{PDF and CDF showing reliability using the various SNR values at 1Mbps}\label{fig:pdfsync}	
\end{figure}

Intuitively higher SNR values produce less transmission errors which allows for higher baud rates to be achieved. After an SNR of approximately 18dB there is no significant changes stemming from increasing the baud rate. Better results are then achieved by adjusting the comparator reference voltage. Figure \ref{fig:pdf1M} and Figure \ref{fig:cdf1M} shows the substitution error frequency and reliability for the results of varying SNR while keeping all other parameters constant. From these figures it is apparent that there are few effects achieved by increasing the SNR beyond 18dB. However, at the lower SNR values the SNR heavily affects the dropped packets as well as error probability and can be mathematically shown in Equation \eqref{eqn:bererfc} in Appendix \ref{app:probs}.

 \subsection{Effects of Comparator Compensation}
 The effects of increasing the baud rate for a certain SNR effectively reduces the duty cycle of the final signal output of the comparator which is fed into the TTL cable. This, expectedly, causes errors as sampling points are incorrectly inferred or missed all together. By adjusting the comparator value to trigger at a lower reference voltage, the overall duty cycle at the output can effectively be compensated for and essentially provide a much cleaner signal which will be able to trigger the correct sampling by the TTL protocol. Table \ref{table:normal} shows the unadjusted duty cycles when using an optimal voltage reference for the comparator while Table \ref{table:adjusted} shows the various duty cycles when using an altered reference voltage. From the two tables, using a reference voltage lower than that of the optimal reference voltage, produces duty cycles which are closer to 50\% and thus allows for better data transmission. The improvement allows transmission from a once impossible communication state to a state which only requires an FEC and reduces the complexity of FEC for the states that previously required it. 

\begin{table}[]
\centering
\caption{Results at 13.88dB using optimal comparator reference voltage}
\label{table:normal}
\resizebox{\textwidth}{!}{%
\begin{tabular}{|c|c|c|c|c|c|c|c|c|c|c|}
\hline
\textbf{\begin{tabular}[c]{@{}c@{}}Baud Rate\\ (Kbps)\end{tabular}} & \textbf{\begin{tabular}[c]{@{}c@{}}TIA\\ + Duty\end{tabular}} & \textbf{\begin{tabular}[c]{@{}c@{}}TIA\\ - Duty\end{tabular}} & \textbf{\begin{tabular}[c]{@{}c@{}}Amp\\ + Duty\end{tabular}} & \textbf{\begin{tabular}[c]{@{}c@{}}Amp\\ - Duty\end{tabular}} & \textbf{\begin{tabular}[c]{@{}c@{}}Amp\\ Min\\ (V)\end{tabular}} & \textbf{\begin{tabular}[c]{@{}c@{}}Amp\\ Max\\ (V)\end{tabular}} & \textbf{\begin{tabular}[c]{@{}c@{}}Ideal\\ Comparator\\ Reference\\ (V)\end{tabular}} & \textbf{\begin{tabular}[c]{@{}c@{}}Actual\\ Comparator\\ Reference\\ (V)\end{tabular}} & \textbf{\begin{tabular}[c]{@{}c@{}}Comparator\\ + Duty\end{tabular}} & \textbf{\begin{tabular}[c]{@{}c@{}}Comparator\\ - Duty\end{tabular}} \\ \hline
\textbf{500}              & 32.5                & 67.5                & 30                  & 70                  & 0.72                 & 3.28                 & 2                                       & 2.01                                     & 70                        & 30                       \\ \hline
\textbf{750}              & 30.77               & 69.23               & 24.2                & 75.8                & 0.72                 & 3.28                 & 2                                       & 2.01                                     & 74.5                      & 25.5                     \\ \hline
\textbf{1000}             & 11                  & 89                  & 20                  & 80                  & 0.72                 & 3.28                 & 2                                       & 2.01                                     & 79                        & 21                       \\ \hline
\end{tabular}%
}
\end{table}

\begin{table}[]
\centering
\caption{Results at 13.88dB using adaptive comparator reference voltage}
\label{table:adjusted}
\resizebox{\textwidth}{!}{%
\begin{tabular}{|c|c|c|c|c|c|c|c|c|c|c|}
\hline
\textbf{\begin{tabular}[c]{@{}c@{}}Baud Rate\\ (Kbps)\end{tabular}} & \textbf{\begin{tabular}[c]{@{}c@{}}TIA\\ + Duty\end{tabular}} & \textbf{\begin{tabular}[c]{@{}c@{}}TIA\\ - Duty\end{tabular}} & \textbf{\begin{tabular}[c]{@{}c@{}}Amp\\ + Duty\end{tabular}} & \textbf{\begin{tabular}[c]{@{}c@{}}Amp\\ - Duty\end{tabular}} & \textbf{\begin{tabular}[c]{@{}c@{}}Amp\\ Min\\ (V)\end{tabular}} & \textbf{\begin{tabular}[c]{@{}c@{}}Amp\\ Max\\ (V)\end{tabular}} & \textbf{\begin{tabular}[c]{@{}c@{}}Ideal\\ Comparator\\ Reference\\ (V)\end{tabular}} & \textbf{\begin{tabular}[c]{@{}c@{}}Actual\\ Comparator\\ Reference\\ (V)\end{tabular}} & \textbf{\begin{tabular}[c]{@{}c@{}}Comparator\\ + Duty\end{tabular}} & \textbf{\begin{tabular}[c]{@{}c@{}}Comparator\\ - Duty\end{tabular}} \\ \hline
\textbf{500}              & 32.5                & 67.5                & 30                  & 70                  & 0.72                 & 3.28                 & 2                                       & 0.881                                    & 56                        & 44                       \\ \hline
\textbf{750}              & 30.77               & 69.23               & 24.2                & 75.8                & 0.72                 & 3.28                 & 2                                       & 0.881                                    & 54.5                      & 45.5                     \\ \hline
\textbf{1000}             & 11                  & 89                  & 20                  & 80                  & 0.72                 & 3.28                 & 2                                       & 0.881                                    & 48                        & 52                       \\ \hline
\end{tabular}
}
\end{table}
\section{Practical Application Making use of TTL OOK VLC system for Media Streaming }
As an extension to the results and to prove the practicality of such a system, the VLC set up which makes use of the TTL protocol and OOK is easily adapted to transmit video files over the visible light spectrum.

    \begin{figure}[hb]
	\centering
		\centering
		\includegraphics[width=8cm]{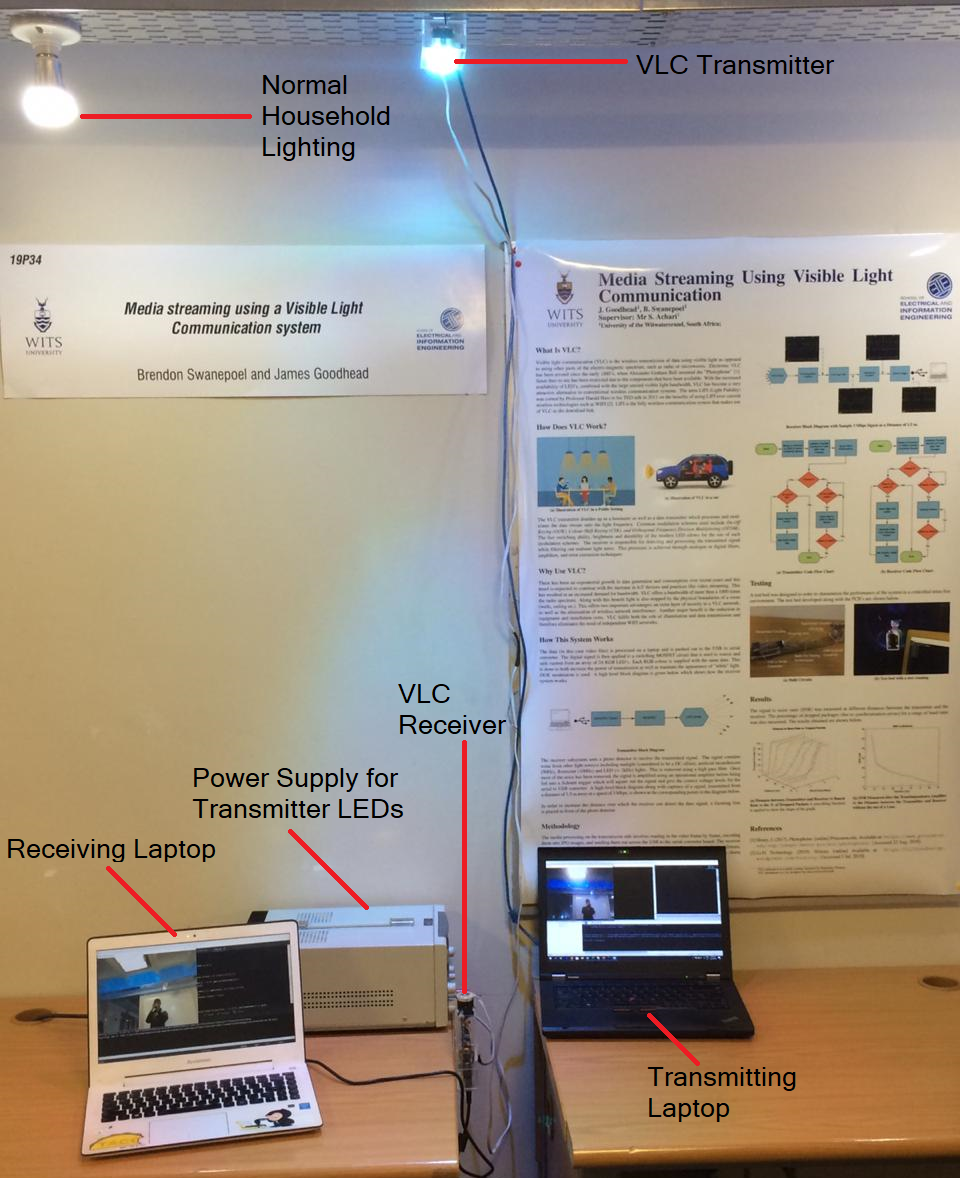}
		\caption{Practical demonstration of the VLC system used to transmit a live video stream}\label{fig:vsapp}		
	\end{figure}

The transmitter and receiver circuit layout is similar to what was presented earlier with the only major differences being that a 12Mbps USB-TTL cable is utilised instead of a 3Mbps cable; the comparator is changed to a Schmitt trigger to allow a smoother turn on/ off range;  lastly, a capacitor is added in series after the TIA to remove the DC offset.  These minor adjustments in hardware immediately reap results as faster baud rates are now possible which in turn leads to faster data communication. This once again reinforces the fact that how the data is synchronised is the novel idea in this paper, as faster speeds are easily attained with trivial modifications to the hardware.
The media streaming system can reliably transmit video data at up to 4Mbps without the use of any error correction schemes and the complete system which is used for everyday application is shown in Figure \ref{fig:vsapp}. Here a live stream from the transmitting laptop is transmitted over VLC and the captured data is shown on the receiving laptop.

\begin{figure}[H]	
	\centering
	\begin{subfigure}[]{5cm}
		\centering
		\includegraphics[width=3cm]{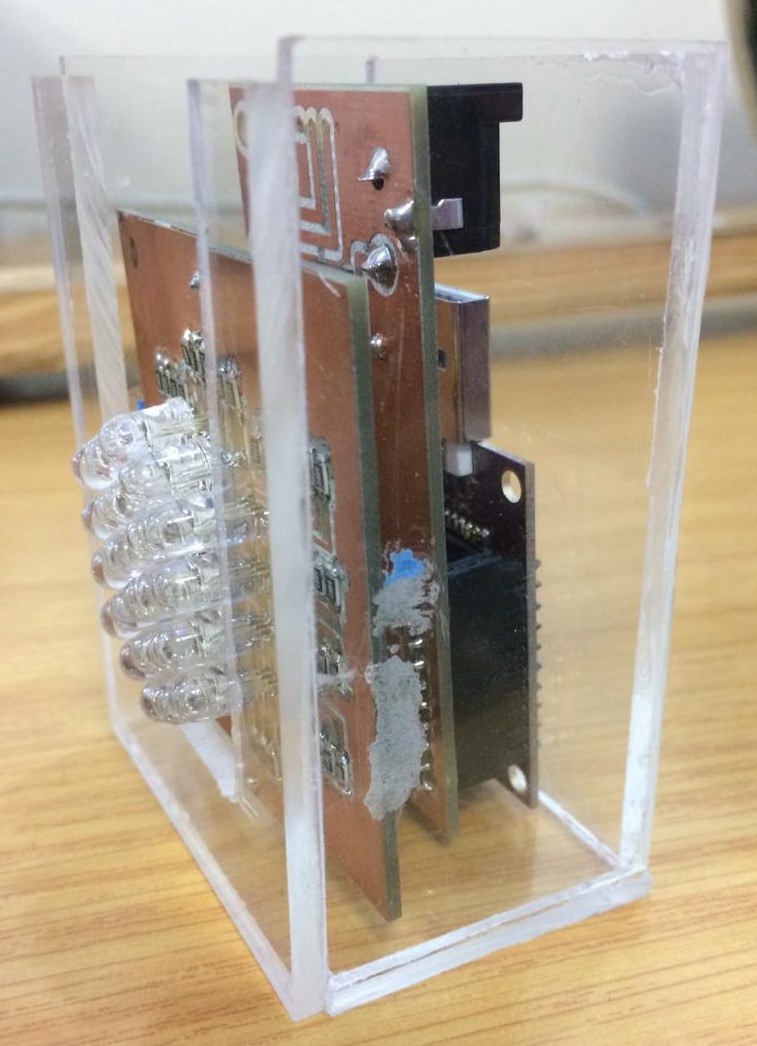}
		\caption{VLC Video streaming transmitter}\label{fig:vstx}		
	\end{subfigure}
	\quad
    	\begin{subfigure}[]{5cm}
		\centering
		\includegraphics[width=5cm]{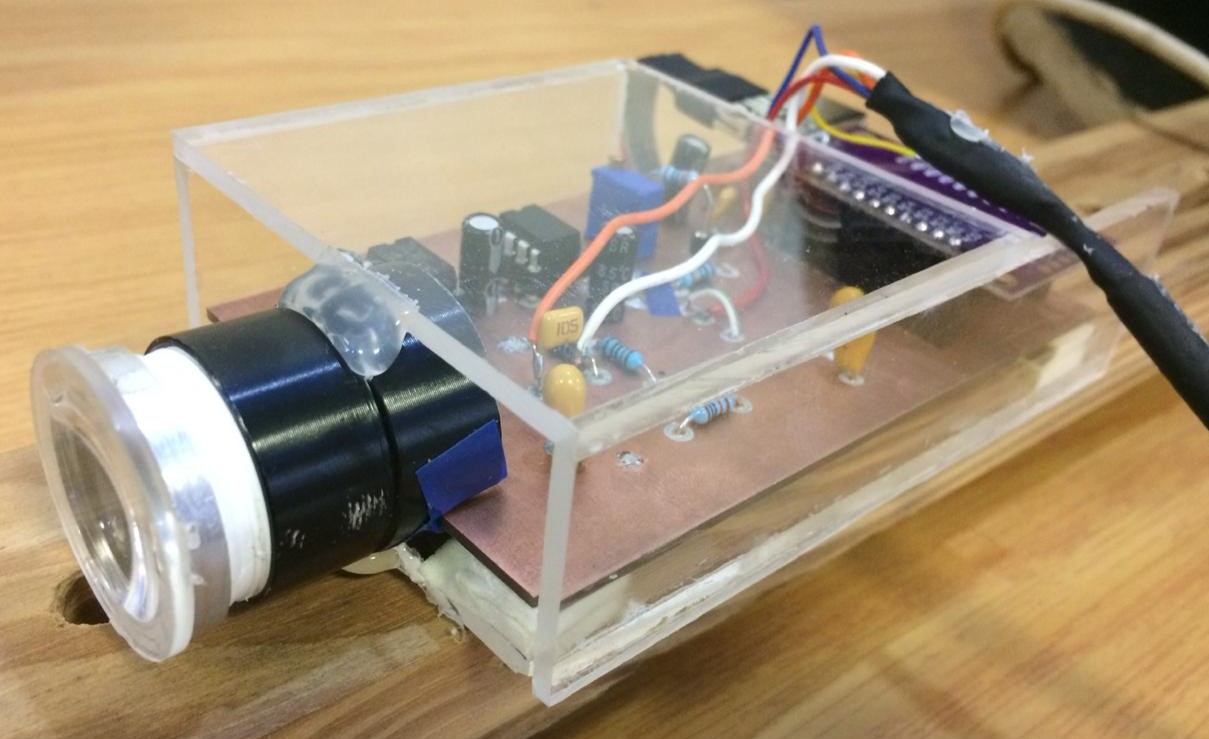}
		\caption{VLC Video streaming receiver}\label{fig:vsrx}
	\end{subfigure}
	\caption{Prototyped circuits of VLC video streaming system}\label{fig:vscct}	
\end{figure}

\begin{figure}[H]	
	\centering
    	\begin{subfigure}[]{10cm}
		\centering
		\includegraphics[width=9cm]{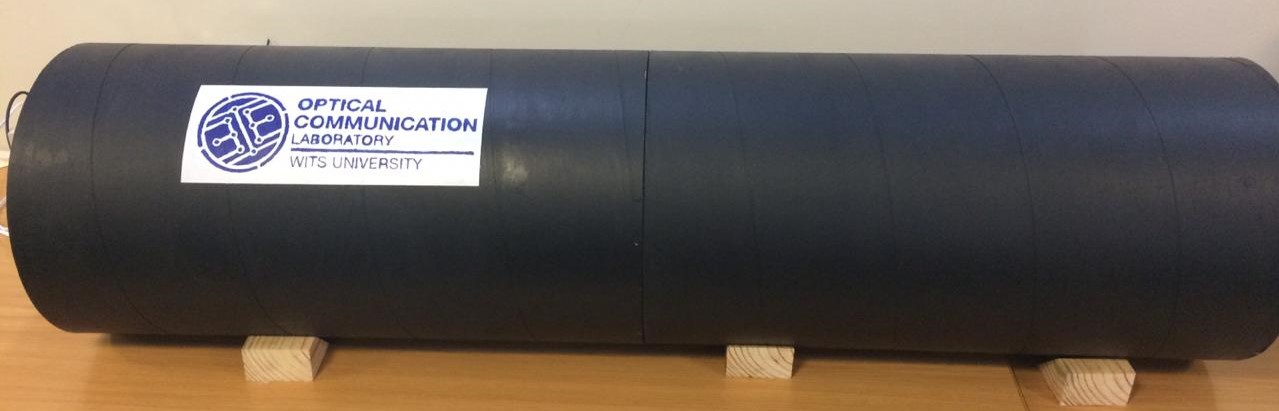}
		\caption{Exterior of video streaming testbed}\label{fig:tb4}
	\end{subfigure}
		\begin{subfigure}[]{5cm}
		\centering
		\includegraphics[width=3.5cm]{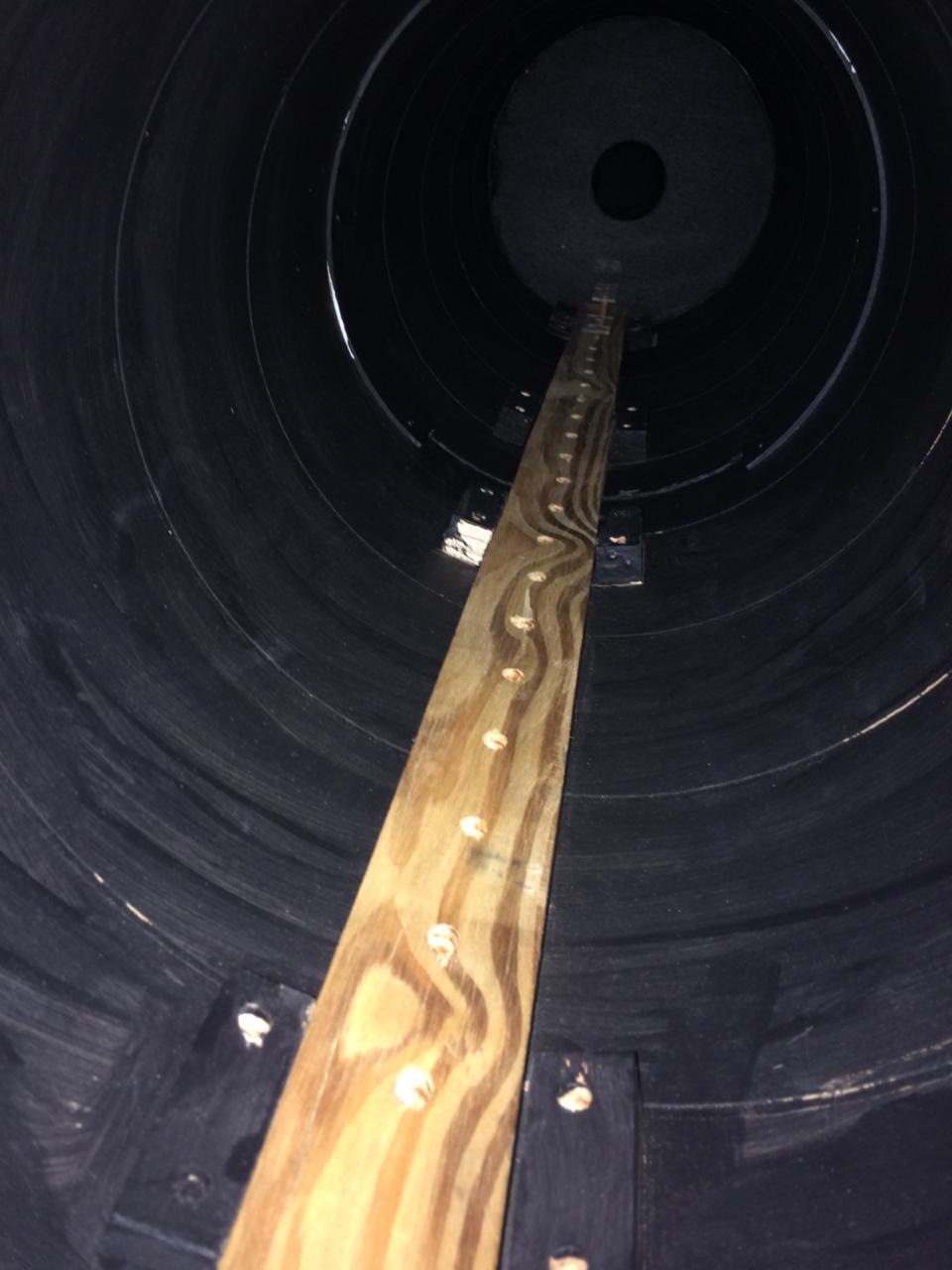}
		\caption{Interior of video streaming testbed}\label{fig:tb3}		
	\end{subfigure}
    	\begin{subfigure}[]{5cm}
		\centering
		\includegraphics[width=3.5cm]{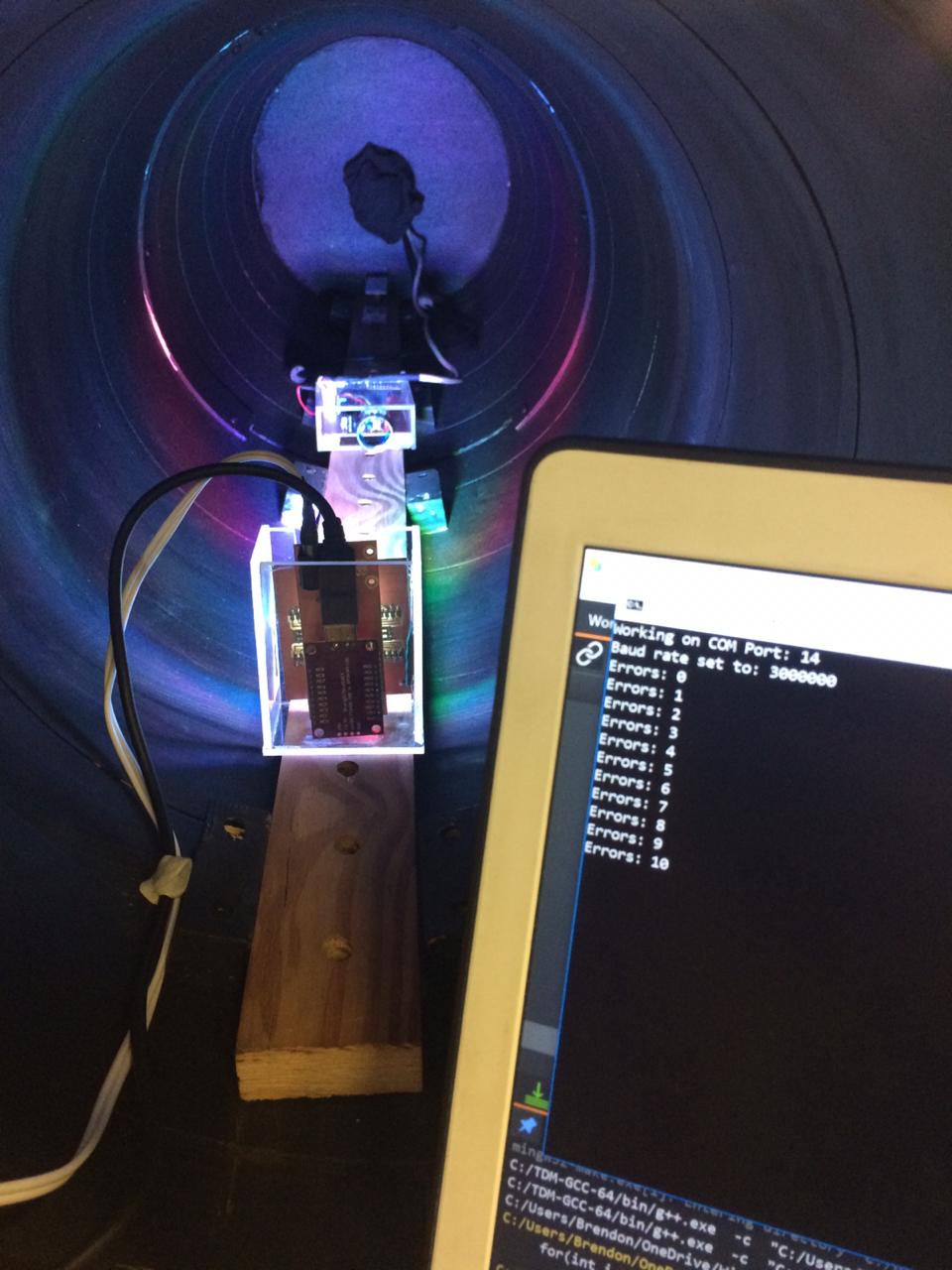}
		\caption{Sample test run within the test bed}\label{fig:tb7}
	\end{subfigure}
	\caption{Test bed and testing of video streaming with VLC}\label{fig:vstb}	
\end{figure}

\begin{figure}[H]
	\centering
		\centering
		\includegraphics[width=15cm]{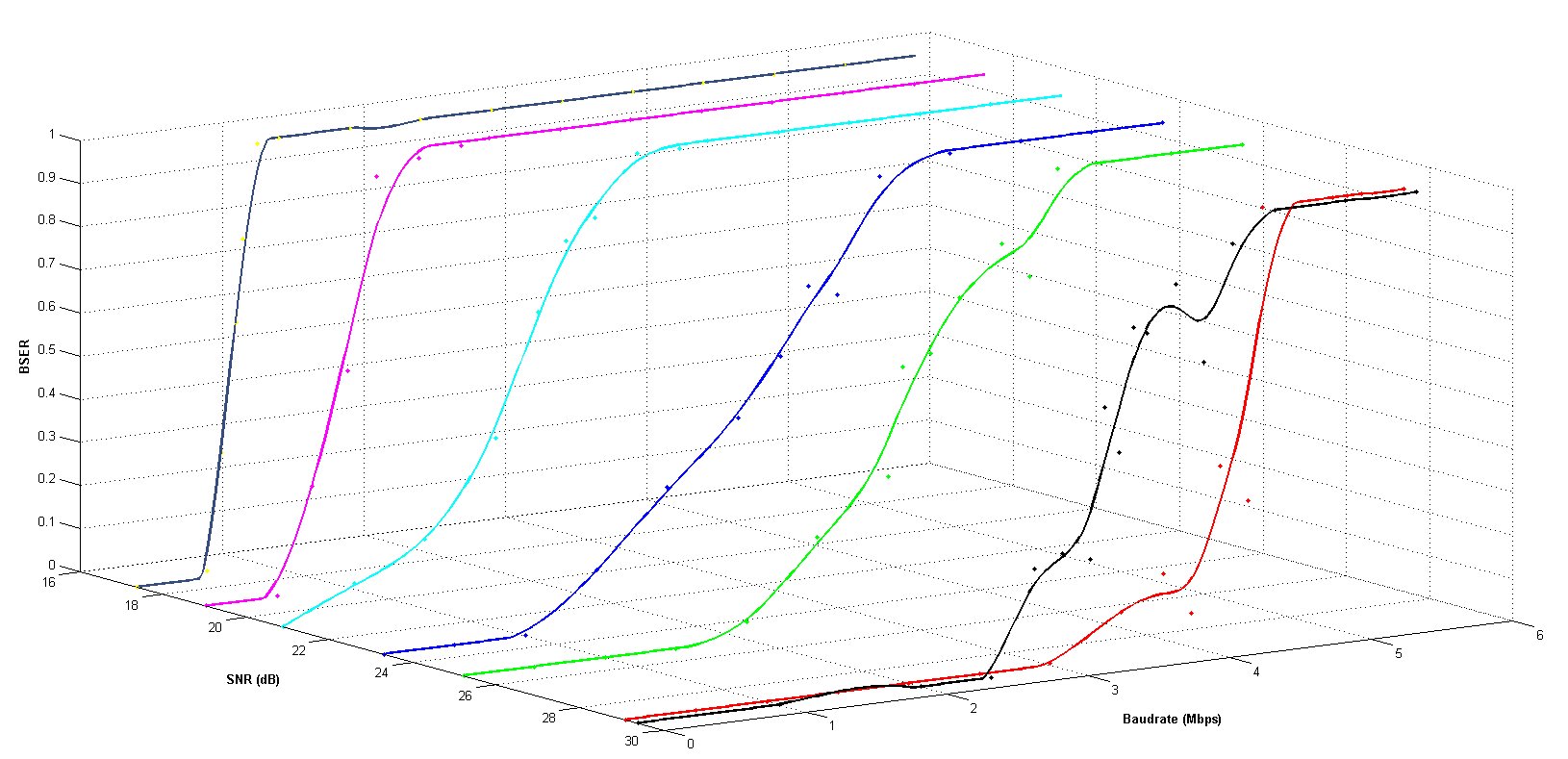}
		\caption{$p_{BSE}$ for various SNR and baud rates with video streaming system}\label{fig:vs3d}		
	\end{figure}

\begin{figure}[H]
	\centering
		\centering
		\includegraphics[width=7cm]{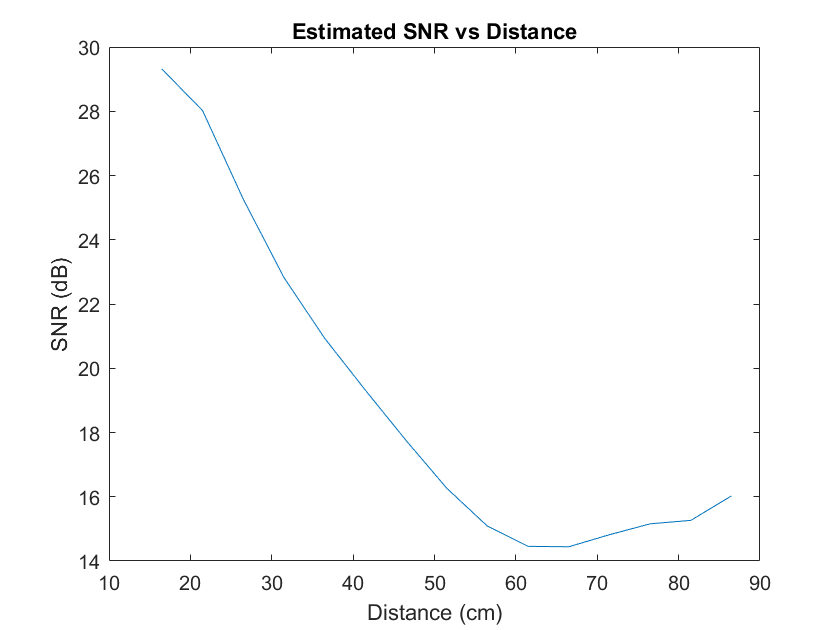}
		\caption{SNR values vs Distance between receiver and transmitter within the test bed}\label{fig:snrvsdis}		
	\end{figure}

Figure \ref{fig:vstx} and Figure \ref{fig:vsrx} show the final transmitter and receiver circuit used for the video streaming.The testbed exterior, testbed interior (with set spacings for the transmitter and receiver) and sample test operation can be seen in Figure \ref{fig:tb4}, Figure \ref{fig:tb3} and Figure \ref{fig:tb7} respectively.
The main results from the video streaming tests and simulations are shown in Figure \ref{fig:vs3d} where the effects of SNR and baudrate on the BSER can once again be observed. It can easily be seen that the trends follow previous tests of the basic system and the analysis is performed similarly as before. The main differences are the higher baud rates achieved. Figure \ref{fig:snrvsdis} additionally shows how SNR can be converted to the distance between the transmitter and receiver within the testbed (ideal conditions). It is worth noting that the lens on the receiver is removed for all tests and the addition of this can ea silly extend the distance between the transmitter and receiver to around 2m in the open air real life application scenario.

\section{Conclusion}
\label{sec:conclusion}
It is possible to achieve reliable communication using a self synchronising inta-vehicle VLC system based on the TTL protocol and OOK modulation within certain regions and parameters. The system is small enough to be incorporated seamlessly into existing vehicle layouts while utilising existing lighting sources. If transmission is not perfect there are FEC coding schemes which may be used to improve the reliability of data transmission. These FEC schemes are for substitution errors and can be inferred from the frequency error distributions and reliability plots. For synchronisation, a simple ARQ protocol is opted for. \color{black} A complete VLC transceiver is built and tested which allows for operating speeds in the range of 10Kbps to 1Mbps and SNR regions from 1.32 dB upwards. This is then improved upon for a practical video streaming application where up to 4Mbps data transmission is achieved without the use of error correction techniques. \color{black} Theoretical probabilities and experimental data are in agreement for tests which show the effect of synchronisation word length, frame length, SNR and comparator reference voltage adjustment.

\appendix

\setcounter{figure}{0}
\setcounter{equation}{0}

\section{Signal-to-Noise Ratio}\label{app:snr}
\renewcommand{\theequation}{\thesection.\arabic{equation}}
\renewcommand{\thefigure}{\thesection.\arabic{figure}}

The SNR values provides a useful metric to determine the reliability of a communications system. The following section details the calculations used for the SNR values for the VLC system using a TTL protocol.

Firstly the following calculations are approximations as no Analog-to-Digital Converter (ADC) is used and consequently no sampling takes place to produce a set of values.

\begin{figure}[h]
	\centering
		\centering
		\includegraphics[width=5.5cm]{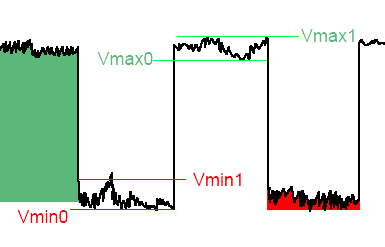}
		\caption{Square wave showing points for SNR calculation}\label{fig:snr2}		
	\end{figure}

SNR is the ratio of signal power compared to that of noise power. From Figure \ref{fig:snr2} there are two easily observable areas which are shaded in green and red. The green shaded area shows the area where signal and noise are present, $P_{SN}$, while the red shaded area shows noise only, $P_N$.

Since only the voltage is observable, it is converted to power by squaring and using a resistance of unity. Also the noise acts as an erratic random wave and is essentially modelled as a sine wave to reduce complexity.  From these points, equations for $P_N$ and $P_{SN}$ are derived and are shown in Equation \eqref{eqn:pn} and Equation \eqref{eqn:psn}, respectively, where the values for $V_{min0}$, $V_{max0}$, $V_{min1}$ and $V_{max1}$ are shown in Figure \ref{fig:snr2}.

\begin{equation}
\label{eqn:pn}
	P_N = (V_{min0})^2 + \left(\frac{V_{min1}-V_{min0}}{2}\right)^2/2
\end{equation}

\begin{equation}
\label{eqn:psn}
	P_{SN} = (V_{max0})^2 + \left(\frac{V_{max1}-V_{max0}}{2}\right)^2/2
\end{equation}

By subtracting Equation \eqref{eqn:pn} from Equation \eqref{eqn:psn} an equation for $P_S$ is derived. Using $P_S$, an equation for SNR is calculated to be $\frac{P_S}{2P_N}$. The coefficient of $\frac{1}{2}$ is due to the randomisation of bits sent resulting in a 50\% chance for a one and 50\% chance for a zero being transmitted.

\subsection{Line of Sight and Non-Line of Sight Theory}
The Line of Sight (LOS) and Non-Line of Sight (NLOS) signals play a huge role in the transmission of data especially in a VLC system. Theoretical values of SNR may additionally be obtained by making use of Equation \eqref{eqn:los} and Equation \eqref{eqn:nlos} which describes the channel gain for both LOS and NLOS links respectively  \cite{kahn1997wireless}. The received power can easily be obtained by multiplying this channel gain by the transmit power.
    \begin{figure}[h]
	\centering
		\centering
		\includegraphics[width=7cm]{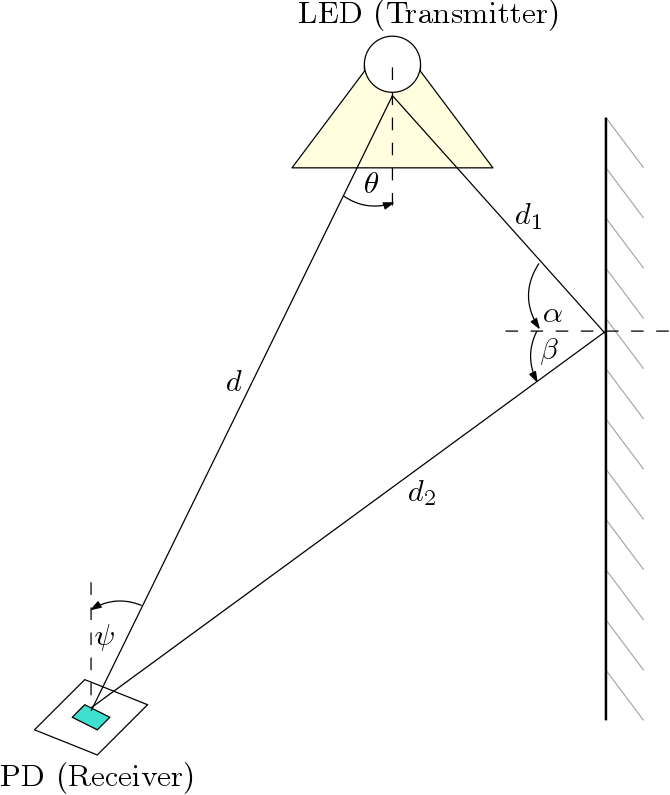}
		\caption{LOS and NLOS links when transmitting over VLC channel for a first order reflection scenario}\label{fig:losnlos}		
	\end{figure}

\newpage
\begin{equation}
\label{eqn:los}
H_{LOS} = 
\begin{cases} 
      \frac{A_r(m+1)}{2 \pi d^2} cos^m(\theta)cos(\psi) & 0 \leq \psi \leq \Psi_c \\
      0 & \text{elsewhere} \\
   \end{cases}
\end{equation}
Where:
\begin{tabular}{llll}
$A_r$  & = Effective area of collection of the photodiode \\
$m$  & = Lambertian emission order (shown in Equation \eqref{eqn:lambertian})  \\ 
$\theta$  & = Angle of irradiance\\
$\psi$ & = Angle of incidence\\
$d$ & = Distance between the transmitter and receiver\\
\end{tabular}

\begin{equation}
\label{eqn:nlos}
	\begin{aligned}
		H_{NLOS} = \frac{A_r(m+1)}{2(\pi d_1 d_2)^2} cos^m(\theta)cos(\alpha)cos(\beta)cos(\psi)d_{A} \rho
	\end{aligned}
\end{equation}
Where:
\begin{tabular}{llll}
$A_r$  & = Effective area of collection of the photodiode \\
$m$  & = Lambertian emission order: calculated using Equation \eqref{eqn:lambertian})  \\ 
$\theta$  & = Angle of irradiance\\
$\psi$ & = Angle of incidence\\
$\alpha$  & = Angle of irradiance wrt normal of reflection\\
$\beta$ & = Angle of incidence wrt normal of reflection\\
$d_1$ & = Distance between the transmitter and reflective surface\\
$d_2$ & = Distance between the reflective surface and receiver\\
$d_{A}$ & = Effective area of reflective surface \\
$\rho$ & = Reflection coefficient of surface \\
\end{tabular}

\begin{equation}
\label{eqn:lambertian}
	\begin{aligned}
		m = \frac{-\ln 2}{\ln cos(\Phi_{1/2})}
	\end{aligned}
\end{equation}
Where:
\begin{tabular}{llll}
$\Phi_{1/2}$  & = Transmitter half angle \\
\end{tabular}

\setcounter{figure}{0}
\setcounter{equation}{0}
\section{Probabilities and BER calculations}\label{app:probs}

\subsection{Bit Error Rate and Symbol Error Rate}
This section focuses on the Bit Error Rate (BER) for a Non Return to Zero On Off Keying (NRZ-OOK) modulation scheme. A 1 is effectively transmitted at a certain total power while a 0 is transmitted at a fraction (usually zero) of this total power. The BER can be expressed in terms of its decibel (dB) SNR and the complementary error function as well as in terms of the Q function shown in Equation \eqref{eqn:bererfc}.

\begin{equation}
\label{eqn:bererfc}
    p_{b} = \frac{1}{2} erfc \ \left(\ {\sqrt[]{\frac{SNR}{2}}}\right) = Q(\sqrt[]{SNR})
\end{equation}

Since the protocol used for transmission of data is the TTL protocol, the SER is achieved by multiplying the data bits sent by the BER. Equation \ref{eqn:sererfc} shows the SER using the complementary error function and the Q function. The main cause of the SER is due to the start and stop bits of the TTL protocol. If either of  these bits are not sampled correctly, it results in either a dropped symbol or propagated error till synchronisation is regained at the protocol level.

\begin{equation}
\label{eqn:sererfc}
	SER_{TTL} = 4 \ erfc\ \left(\ {\sqrt[]{\frac{SNR}{2}}}\right) = 8 \ Q(\sqrt[]{SNR})
\end{equation}

\subsection{Probabilities}
The two main probabilities of interest at the frame level are that of synchronisation failure and synchronisation error denoted $p_{fail}$ and $p_{err}$, respectively. Synchronisation failure occurs when the packet or frame is dropped. This happens when the received frame length is not equal to the sent frame length and occurs in one of two ways. First, when any symbol from the synchronisation word is lost or converted to another letter in the alphabet, or second when a symbol from the payload is converted into the synchronisation symbol. Let \textit{$n_{sync}$} denote the number of synchronisation symbols in the synchronisation word, \textit{$n_{payload}$} be the number of symbols in the payload of the frame and \textit{$p_s$} be the crossover probability (or SER). Equation \eqref{eqn:pfail} shows this failure probability. Since the alphabet is ASCII based there are eight bits which provide 256 unique symbols and characters. This provides the coefficient in Equation \eqref{eqn:pfail} and Equation \eqref{eqn:perr} as the probability of going from any symbol in the payload to the synchronisation symbol is $\frac{1}{256-1}$.

\begin{equation}
\label{eqn:pfail}
	p_{fail} = n_{sync} \ p_s + n_{payload}\left(\frac{p_s}{255}\right)
\end{equation}

Synchronisation error is a much less frequent occurrence, as this is the probability that the frame length is as required. However, the synchronisation words are moved, thus resulting in substitution errors within the frame. For this to occur, two sets of synchronisation words need to be converted to other characters of the alphabet. In addition, two sets of the payload need to be converted into synchronisation symbols and also be the correct distance apart from each other. Equation \eqref{eqn:perr} shows this error probability.

\begin{equation}
\label{eqn:perr}
	p_{err} = n_{payload} \ p_s^{\ 2 n_{sync}} \ \left(\frac{p_s}{255}\right)^{2 n_{sync}}
\end{equation}

\section*{acknowledgements}
This work is based upon research supported by the South African National Research Foundation (Grant No. 112248).



\bibliography{sample}



\end{document}